\documentclass[12pt]{article}
\usepackage{amssymb, amsmath, amsthm}
\usepackage{graphicx}
\usepackage{fullpage,booktabs}
\usepackage{siunitx}
\usepackage{subcaption}
\usepackage{multirow}
\newcommand{\argmax}{\operatornamewithlimits{arg\,max}}
\newcommand{\argmin}{\operatornamewithlimits{arg\,min}}

\usepackage{mathtools}
\usepackage{lipsum}
\newcommand{\minim}{\operatorname*{minimize}}

\newcommand{\mbX}{\mathbf{X}}

\newcommand{\mbB}{\boldsymbol{B}}

\newcommand{\mbe}{\mathbf{e}}

\newcommand{\mbS}{\boldsymbol{S}}
\newcommand{\mbs}{\boldsymbol{s}}
\newcommand{\mbD}{\boldsymbol{D}}

\newcommand{\mbb}{\boldsymbol{b}}

\newcommand{\mbW}{\boldsymbol{W}}

\newcommand{\mbZ}{\mathbf{Z}}
\newcommand{\mbv}{\boldsymbol{v}}
\newcommand{\mbr}{\boldsymbol{r}}
\newcommand{\mbu}{\boldsymbol{u}}
\newcommand{\mbt}{\boldsymbol{t}}

\newcommand{\mbz}{\boldsymbol{z}}
\newcommand{\mby}{\boldsymbol y}

\newcommand{\mbSigma}{\boldsymbol{\Sigma}}

\newcommand{\mbbeta}{\boldsymbol{\beta}}
\newcommand{\mbtheta}{\boldsymbol{\theta}}
\newcommand{\mbgamma}{\boldsymbol{\gamma}}

\renewcommand{\hat}{\widehat}

\usepackage{natbib}
\usepackage{fullpage}

\usepackage[colorlinks,citecolor=blue,urlcolor=blue,linkcolor=red,naturalnames=true]{hyperref}
\usepackage{amsmath}
\usepackage[table]{xcolor}
\newcolumntype{L}{>$l<$}
\theoremstyle{plain}

\newtheorem{remark}{Remark}

\usepackage{algorithm,algorithmic}
\usepackage{setspace}
\usepackage{blindtext}

\usepackage{tocloft}
\usepackage{xr}

\begin{document}

\title{\textbf{Smooth and shape-constrained quantile distributed lag models}}
\author{Yisen Jin$^\dagger$, Aaron J. Molstad\footnote{Correspondence: amolstad@umn.edu}~$^{\dagger}$, Ander Wilson$^\ddagger$, and Joseph Antonelli$^\dagger$\\
Department of Statistics, University of Florida, Gainesville, FL$^\dagger$\\
School of Statistics, University of Minnesota, Minneapolis, MN$^*$\\
Department of Statistics, Colorado State University, Fort Collins, CO$^\ddagger$
}
\date{}
\maketitle

\begin{abstract}
  Exposure to environmental pollutants during the gestational period can significantly impact infant health outcomes, such as birth weight and neurological development. Identifying critical windows of susceptibility, which are specific periods during pregnancy when exposure has the most profound effects, is essential for developing targeted interventions. Distributed lag models (DLMs) are widely used in environmental epidemiology to analyze the temporal patterns of exposure and their impact on health outcomes. However, traditional DLMs focus on modeling the conditional mean, which may fail to capture heterogeneity in the relationship between predictors and the outcome. Moreover, when modeling the distribution of health outcomes like gestational birthweight, it is the extreme quantiles that are of most clinical relevance. We introduce two new quantile distributed lag model (QDLM) estimators designed to address the limitations of existing methods by leveraging smoothness and shape constraints, such as unimodality and concavity, to enhance interpretability and efficiency. We apply our QDLM estimators to the Colorado birth cohort data, demonstrating their effectiveness in identifying critical windows of susceptibility and informing public health interventions.
  \end{abstract}
  
\def\spacingset#1{\renewcommand{\baselinestretch}%
{#1}\small\normalsize} \spacingset{1}

\section{Introduction}

The gestational period is a critical developmental stage where exposure to environmental pollutants can profoundly affect birth and children's health outcomes. During pregnancy, the fetus undergoes rapid growth and development, making it particularly vulnerable to environmental influences. Increased exposure to pollutants during pregnancy has been linked to adverse outcomes such as decreased birth weight, increased risk of asthma, preterm birth, and altered neurological development \citep{bosetti2010ambient, stieb2012ambient, jacobs2017association}. Common pollutants include particulate matter, heavy metals, and various organic compounds, which can interfere with normal biological processes. Recent studies have focused on leveraging high-resolution exposure data collected throughout pregnancy to identify critical periods of vulnerability, which are specific times during development when exposure can impact future health outcomes \citep{wright2017environment}. Understanding these so-called ``critical windows'' can help in developing targeted interventions to mitigate the adverse effects of environmental exposures.

The distributed lag model (DLM) is widely used in environmental epidemiology for modeling the relationship between health outcomes and time-dependent exposure to environmental pollutants. A DLM regresses the outcome on repeated measures of exposures over a preceding time period, capturing the temporal pattern of exposure and its impact on health. This model is particularly useful for understanding how exposures at different times during pregnancy affect health outcomes such as birth weight  \citep{schwartz2000distributed,zanobetti2000generalized}. Typically, DLMs are constrained so that the exposure effects vary smoothly over time to reduce the effects of multicollinearity in the repeated measures of exposures. Compared to using exposures averaged over a prespecified time window, such as trimester average exposures, a constrained DLM can reduce estimation error and improve recovery of critical windows \citep{wilson2017potential}. Some recent work on distributed lag modeling has focused on the estimation of nonlinear exposure-time-response functions, allowing for more flexible modeling of complex relationships \citep{gasparrini2014modeling,gasparrini2017penalized,mork2023estimating}. A recent focus in environmental health research has been on mixtures of exposures, or simultaneous exposure to multiple pollutants \citep{joubert2022powering}. There are several recently proposed methods for identifying critical windows to a mixture \citep{warren2022critical,wilson2022kernel,mork2023estimating,antonelli2024multiple}. All of these methods consider mean regression.

While existing methods can work well for modeling the conditional expectation of the outcome, in environmental epidemiology, investigators are often interested in modeling specific quantiles of the outcome as a function of covariates.  Moreover, outcomes are often heteroscedastic and can exhibit skewness, in which case least squares-based estimators of the conditional mean may perform poorly. 
  Quantile regression addresses these issues by allowing one to model conditional quantiles of the outcome, which may provide a more comprehensive perspective of the effects of environmental exposures \citep{koenker1978regression,yu2003quantile}. For instance, quantile regression can reveal how exposure impacts not only the median birth weight but also the lower and upper tails of the birth weight distribution, which are of greatest clinical concern. Moreover, a quantile distributed lag modes (QDLM) may capture the heterogeneity in the effects of exposures over different quantiles of the outcome distribution, providing insights into how exposures might affect individuals differently depending on their position within the outcome distribution \citep{wang2023semiparametric}. This makes quantile regression a powerful tool for environmental health studies, where understanding the full range of potential impacts is crucial. 
  
In this paper, we introduce two new estimators of the QDLM. These estimators exploit common assumptions about distributed lag models: namely, that effects are smooth across time. To enhance interpretability, we also consider more restrictive shape constraints that are plausible in environmental applications. As we describe in a later section, these shape constraints are especially natural in our birth weight modeling application, but difficult to impose in a QDLM. We use the aforementioned modeling approaches to estimate the time varying health effects of multiple environmental exposures on gestational birth weight in a cohort of births in Colorado, USA. Quantile regression is particularly useful for this application as it is the extremes of the outcome---particularly the lower quantiles---that are of utmost importance due to the negative consequences of very low gestational birth weight. Existing methods designed for mean regression are thus insufficient for modeling how exposures affect the lower quantiles of gestational birth weight. 

\section{Quantile distributed lag model}
We focus on modeling the conditional quantiles of the response variable $Y$ as a function of both time-dependent and time-invariant covariates. Specifically, for a given quantile $\tau\in(0,1)$, our goal is to model the $\tau$-th conditional quantile, $q_\tau$, of $Y$ given the entire observed process of $K$ time-dependent predictors $\mbX(\cdot) = (X_1(\cdot), \dots X_k(\cdot))^\top \in \mathbb{R}^K$ at $T$ discrete time points, and $\mbZ \in \mathbb{R}^p$. By definition, ${\rm Pr}\{Y \leq q_\tau(Y\mid \mbX(\cdot), \mbZ) \mid \mbX(\cdot), \mbZ\} = \tau.$ We assume the QDLM 
\begin{equation}\label{eq:model_complex}
  q_\tau(Y\mid \mbX(\cdot), \mbZ) = \sum_{k=1}^{K}  \sum_{m = 1}^T X_k(t_m) \beta_{*k}(t_m)  + \mbZ^\top\mbgamma_*,
\end{equation}
where $X_k(t_m)$ is the value of the $k$th exposure at time $t_m$ and $\beta_{*k}(t_m)$ is the regression coefficient for the $k$th exposure at time $t_m$ for $k \in [K]$ and $m \in [T]$. Additionally, $\mbgamma_* \in \mathbb{R}^p$ are the regression coefficients for the time-invariant covariates $\mbZ$. We assume that each exposure is measured at the same time points $t_1, \dots, t_T$ to simplify our descriptions: we discuss how our method can handle distinct time points in a later section.


To simplify the notation, redefine $\mbX \in \mathbb{R}^{K \times T}$ and $\mbbeta_* \in \mathbb{R}^{K \times T}$ be a matrices with $(k,m)$th entries $X_k(t_m)$ and $\beta_{*k}(t_m)$, respectively. With this notation, we can write \eqref{eq:model_complex} more compactly as 
\begin{equation}\label{eq:model_simple}
  q_\tau(Y\mid \mbX, \mbZ) = {\rm tr}(\mbX^\top \mbbeta_*)  + \mbZ^\top\mbgamma_*,
\end{equation}
where ${\rm tr}(\cdot)$ is the operator that sums the diagonal elements of its matrix-valued argument. 
Denote the $k$th row of $\mbbeta_{*}$ as $\mbbeta_{*k} = (\beta_{*k}(t_1), \dots, \beta_{*k}(t_T))^\top \in \mathbb{R}^T$ where $\beta_{*k}(t_m)$ is the effect of the $k$th exposure at time $t_m$ on the $\tau$th quantile of the response.  For simplicity of notation, the dependence of the $\mbbeta_{*}$ and $\mbgamma_*$ on $\tau$ is omitted.

The goal is to estimate the unknown parameters $\mbbeta_{*}$ and $\mbgamma_*$. In our motivating application---modeling the effect of pollutants on birthweight---the signal to noise ratio is often low and the correlation among exposures is very high. In such settings, it is often beneficial to impose structural constraints on the estimated coefficients in order to improve efficiency and interpretability. For example, in previous work on quantile distributed lag modeling, time-dependent effects were estimated using splines to achieve smoothness of the estimated coefficients over time. Specifically, \citet{wang2023semiparametric} approximated \(\sum_{k=1}^{K}\sum_{m=1}^{T}\beta_{*k}(t_m)X_{k}(t_m)\) using B-splines. 

In this work, we estimate the coefficients $\mbbeta_*$ directly.  Our estimator is motivated partly from a functional data perspective. Consider, for a moment, the functional version of our (discrete) QDLM given by
$$  q^f_\tau(Y\mid \mbX(\cdot), \mbZ) = \sum_{k=1}^{K}   \int_0^V X_k(t) \beta^f_{*k}(t) dt  + \mbZ^\top\mbgamma_*,$$
where here, $\beta^f_{*k}(t)$ is a smooth function over time $t \in [0,V].$ Often, when estimating functions like $\beta^f_*$, it is natural to impose shape constraints, such as concavity (convexity) or unimodality \citep{ghosal2023shape}. In the context of our motivating data analysis, the concavity of $\beta^f_{*k}(t)$ is especially natural, as it indicates that the $k$th exposure's effect gradually increases, peaks at some point during pregnancy, then gradually decreases in a structured way. Unimodality is less restrictive than concavity, requiring only a single extremum for $\beta^f_{*k}(\cdot)$ and monotonicity about this extremum. If the extremum is a maximum, it can be interpreted similarly: the $k$th exposure's effect first increases, peaks mid-pregnancy, then decreases. However, the changes in effects before and after the peak are less restrictive compared to concavity. These shape constraints enhance interpretability. For example, if the function $\beta^f_{*k}(\cdot)$ is concave or unimodal, then we can infer that there is a time period in which exposure to a pollutant is most harmful. In contrast, if $\beta^f_{*k}(t)$ fluctuates across time $t$, it becomes more challenging to describe the effect of pollutant $k$ on the quantile of the outcome.

Borrowing inspiration from the functional model under shape constraints, to fit \eqref{eq:model_simple}, we make two key assumptions. First, we assume that the $\beta_{*k}(\cdot)$ are smooth over time (i.e., $\beta_{*k}(t_m)$ is similar to $\beta_{*k}(t_{m'})$ when $t_m$ is close to $t_{m'}$). This assumption is consistent with the smoothness of $\beta^f_{*k}(t)$ for $t \in [0, V]$. Secondly, we assume that the $\beta_{*k}$ are either (i) unimodal or (ii) concave, or can be well approximated by a unimodal or concave function. To achieve concavity of the $k$th exposure's effect under the discrete model \eqref{eq:model_simple}, with equally spaced time points, it would require that $\beta_{*k}(t_m) + \beta_{*k}(t_{m+2}) \leq 2 \beta_{*k}(t_{m+1})$. For the remainder of this and subsequent sections, we will refer to assumption (ii) as concavity, though convexity can be achieved by flipping the sign of the corresponding exposure. Similarly, when describing a unimodal function, we will assume its extremum is a maximum for ease of exposition. We will propose separate estimators for (i) and (ii) under the smoothness assumption, both of which we describe in the next section.

\section{Smooth and shape-constrained QDLM estimators}
\subsection{Overview}
To understand how we can impose structural constraints in the discrete model, let us first consider the (idealized) case that we know a priori that $M_{*k} = \argmax_{m \in [T]} \mbbeta_{*k}(t_m)$ for each $k \in [K]$, and that $\mbbeta_{*k}(t)$ is unimodal. Here, $M_{*k} \in [T]$ for each $k \in [K]$. 
Suppose we have observed triplets $\{y_i, \mbX_i, \mbZ_i\}_{i=1}^n$ where $y_i \in \mathbb{R}$ is the $i$th subject's observed response, $\mbX_{i} \in \mathbb{R}^{K \times T}$ has $(k,m)$th entry being the $i$th subject's exposure to the $k$th pollutant at time $t_m$, and $\mbZ_i \in \mathbb{R}^p$ is the $i$th subject's covariates.  Let $\rho_\tau(a) = a \{\tau - \mathbf{1}(a < 0)\}$ be the check loss function for the $\tau$th quantile evaluated at $a \in \mathbb{R}.$ Then, to fit a smooth and unimodal distributed lag regression model, we could use the constrained estimator 
\vspace{-5pt}
\begin{equation}\label{eq:oracle}
\argmin_{\boldsymbol{\beta}, \mbgamma}\sum_{i=1}^n\rho_{\tau}\left\{ y_i- {\rm tr}(\mbX_i^\top \mbbeta) - \mbZ_i^\top\boldsymbol\mbgamma\right\} \text{ subject to } \sum_{k=1}^K \|\mbD^{(v)} \boldsymbol{\beta}_k\|_2^2\leq \lambda_0,  
\end{equation}
\vspace{-5pt}
$$  \beta_{k}(t_m)\le\beta_{k}(t_{m+1}) ~\text{for}~ m \in [M_{*k}-1],~ \beta_{k}(t_{m})\ge\beta_{k}(t_{m + 1}) ~\text{for}~ m \in [T]\setminus[M_{*k}-1], ~~k \in [K],\vspace{5pt}$$       
where $\lambda_0 > 0$ is a user specified tuning parameter, $\|\cdot\|_2$ is the Euclidean norm, and $\mbD^{(v)} \in \mathbb{R}^{(T - v) \times T}$ is a discrete difference operator of order $v$ \citep{tibshirani2014adaptive}.  In particular,
$$\mbD^{(1)} = \left( \begin{array}{c c c c c c}
1 & -1 & 0 & \cdots & 0 & 0 \\
0 & 1 & -1 & \cdots & 0 & 0 \\
\vdots \\
0 & 0 & 0 & \cdots & 1 & -1
\end{array}\right), ~~~ \mbD^{(v+1)} =  \mbD^{(1)}_{T-v} \mbD^{(v)}, ~~~ v \geq 2,$$
where $\mbD^{(1)}_{T-v} \in \mathbb{R}^{(T - v - 1) \times (T - v)}$ is the version of $\mbD^{(1)} \in \mathbb{R}^{(T-1) \times T}$ above with $T$ replaced with $T - v$.  Therefore, in the criterion \eqref{eq:oracle}, the tuning parameter $\lambda_0$ controls the smoothness of the $\hat\mbbeta_k$ across the $T$ time points. If $v = 1$, for example, then the solution to \eqref{eq:oracle} will satisfy $\sum_{k=1}^K \sum_{m=1}^{T-1}\{\hat\mbbeta_k(t_m) - \hat\mbbeta_k(t_{m+1})\}^2 \leq \lambda_0$. 

While \eqref{eq:oracle} is well-motivated, there are two challenges to its use in practice. First, the $M_{*1}, \dots, M_{*K}$ will, in general, not be known in practice: these must be estimated from the data. Second, while some $\mbbeta_{*k}$ may have a unimodal shape, we want to allow for violations of unimodality or monotonicity.  To borrow phrasing from \citet{tibshirani2011nearly}, we want to allow for ``nearly" unimodal estimates with the degree of unimodality and monotonicity determined by the data.  To this end, define the function
$|\mbD^{(v)} \mbbeta_{k}|^+ = \sum_{j=1}^{T-v} \max\{\mbbeta_k^\top [\mbD^{(v)}]_{j}, 0\}$
where $[\mbD^{(v)}]_{j} \in \mathbb{R}^T$ is the $j$th row of $\mbD^{(v)}$. For example, 
$|\mbD^{(1)} \mbbeta_{k}|^+ = \sum_{m=1}^{T-1} \max\{\beta_k(t_m) - \beta_k(t_{m+1}), 0\}.$
Define also $|\mbD^{(v)} \mbbeta_{k}|^- = \sum_{j=1}^{T-v} \max\{-\mbbeta_k^\top [\mbD^{(v)}]_{j}, 0\}.$ 
Finally, define 
\begin{equation}\label{h:equation}
h^{(v)}(\mbbeta_k; M_k) = |\mbD^{(v)}_{M_k}\mbbeta_{k,1:M_k}|^+ + |\mbD^{(v)}_{T - M_k + 1}\mbbeta_{k,M_k:T}|^-,
\end{equation}
where $\mbbeta_{k,a:b} = (\beta_k(t_a), \dots, \beta_k(t_b))^\top \in \mathbb{R}^{b-a + 1}.$

\subsection{Nearly-unimodal quantile distributed lag estimator}
If we knew the modes $M_{*1}, \dots, M_{*K}$ a priori, we could rewrite the monotonicity constraints from \eqref{eq:oracle} in terms of the function $h$, defined in \eqref{h:equation}. 
Notice, $\beta_{k}(t_m)\le\beta_{k}(t_{m+1}) ~\text{for}~ m \in [M_{*k} - 1]$ if and only if $|\mbD^{(1)}_{M_{*k}}\mbbeta_{k,1:M_{*k}}|^+ = 0$ . Therefore, \eqref{eq:oracle} is equivalent to 
\begin{equation}\label{eq:oracle_reframe}
\argmin_{\boldsymbol{\beta}, \mbgamma}\sum_{i=1}^n\rho_{\tau}\left\{ y_i- {\rm tr}(\mbX_i^\top \mbbeta) - \mbZ_i^\top\boldsymbol\mbgamma\right\} 
\end{equation}
\vspace{-5pt}
    $$ \text{ subject to } \sum_{k=1}^K \|\mbD^{(v)} \boldsymbol{\beta}_k\|_2^2\leq \lambda_0,~~~ h^{(1)}(\mbbeta_k; M_{*k}) = 0, ~~k \in [K].\vspace{5pt}$$
If we replace $h^{(1)}(\mbbeta_k; M_{*k}) = 0$ with $h^{(1)}(\mbbeta_k; M_{*k}) \leq c$ for a positive $c$, we can allow for unimodality or monotonicity to be violated, so our estimator can be ``nearly-unimodal''. Clearly, if $h^{(1)}(\widehat\mbbeta_k; M_{*k}) > 0$, then unimodality about $\widehat\mbbeta_{k, M_{*k}}$ is violated by $\widehat\mbbeta_k$.

With these ideas in mind, we propose to estimate $\mbbeta_*$ under smoothness and nearly-unimodal constraints by solving a penalized version of \eqref{eq:oracle_reframe}
\begin{equation}\label{eq:nearlyUnimodal}
\argmin_{\boldsymbol{\beta}, \mbgamma, \{M_k\}_{k=1}^K }~ \left[ \frac{1}{n}\sum_{i=1}^n \rho_{\tau}\left\{ y_i- {\rm tr}(\mbX_i^\top \mbbeta) - \mbZ_i^\top\boldsymbol\mbgamma\right\} + \lambda_1 \sum_{k=1}^K  h^{(1)}(\mbbeta_k; M_k) + \lambda_2 \sum_{k=1}^K \|\mbD^{(v)}\mbbeta_{k}\|_2^2 \right]
\end{equation}
for user specified tuning parameters $\lambda_1 \geq 0$ and $\lambda_2 > 0.$ To simplify matters, we will take $v = 2$ in the final term of for the remainder of the article. The criterion \eqref{eq:nearlyUnimodal} differs from \eqref{eq:oracle_reframe} in that both (i) unimodality is not enforced strictly, and (ii) the $M_k \in [T]$ are treated as optimization variables. We call the pair of arguments minimizing \eqref{eq:nearlyUnimodal}  with respect to $(\mbbeta, \mbgamma)$ the nearly-unimodal estimator of $(\mbbeta_*, \mbgamma_*)$. It is important to note for computation that with the $M_k$ fixed, the optimization problem in \eqref{eq:nearlyUnimodal}---with respect to $\mbgamma$ and $\mbbeta$---is convex. With the $\mbgamma$ and $\mbbeta$ fixed, the optimization with respect to the $M_k$ is discontinuous, but can be solved via a simple exhaustive grid search, as we discuss later.  

The two penalties in \eqref{eq:nearlyUnimodal} serve different purposes. For example, with $\lambda_1$ large and $\lambda_2 = 0$, the solution $\hat\mbbeta_k$ must be monotonic on either side of its $\hat{M}_k$-th entry, but there may be large jumps (e.g., $\hat\beta_{k}(t_m) - \hat\beta_{k}(t_{m+1})$ for $m \geq \hat{M}_{k}$ may be arbitrarily large). By taking $\lambda_2 > 0$ sufficiently large, the solution must also be relatively smooth across time, i.e., $|\hat\beta_{k}(t_m) - \hat\beta_{k}(t_{m+1})| < \epsilon$ for all $m \in [T]$ for some $\epsilon > 0$. Conversely, by taking $\lambda_1 = 0$, the solution will be smooth across time, but will not be unimodal.  By selecting both $\lambda_1$ and $\lambda_2$ using cross-validation, we allow the data to determine the degree to which smoothness and unimodality improve the model fit. 

While we have thus far focused on the assumption that $\beta_{*k}(t_1)\le\beta_{*k}(t_2)\le\cdots\le\beta_{*k}(t_{M_k})$ and  $\beta_{*k}(t_{M_k})\ge\cdots\ge\beta_{*k}(t_T)$, our method can easily accommodate the assumption that $\beta_{*k}(t_1)\ge\beta_{*k}(t_2)\ge\cdots\ge\beta_{*k}(t_{M_k})$ and  $\beta_{*k}(t_{M_k})\le\cdots\le\beta_{*k}(t_T)$: this is achieved by simply flipping the sign of the corresponding exposure. In applications with a reasonably small number of exposures $K$, it is feasible to try all possible combinations in cross-validation, though we expect in most applications, prior knowledge will determine which assumptions are reasonable. 

In practice, we use a slightly modified version of \eqref{eq:nearlyUnimodal} to simplify computation. This version of our method replaces $h^{(1)}(\mbbeta_k; M_k)$ with $$\overline{h}^{(1)}(\mbbeta_k; M_k) = |\mbD^{(v)}_{M_k}\mbbeta_{k,1:M_k}|^+ + |\mbD^{(v)}_{T - M_k}\mbbeta_{k,(M_k+1):T}|^-.$$
Compared to $h^{(1)}, \overline{h}^{(1)}$ omits $\mbbeta_{k,M_k}$ from the second term in the penalty. Thus, $M_k$ is either the mode, or $M_k + 1$ is the mode. Practically speaking, this has no effect on our fitted model as when the tuning parameter $\lambda_1$ is large, our estimator still enforces unimodality. As we will discuss in a later section, however, by defining $\overline{h}^{(1)}$ as the sum of two terms that depend on distinct subvectors of $\mbbeta_k$, computation is greatly simplified.

\subsection{Nearly-concave quantile distributed lag estimator}
In settings where it may be appropriate to assume concavity of the $\mbbeta_{*k},$ we can utilize penalties similar to \eqref{eq:nearlyUnimodal}. Recall that $\mbbeta_{*k}$ is concave if $2\beta_{*k}(t_{m+1}) \geq \beta_{*k}(t_{m}) + \beta_{*k}(t_{m+2})$ for all $m \in [T-2]$. Expressed in terms of the functions we defined before, $\mbbeta_{*k}$ is concave if $|\mbD^{(2)}\mbbeta_{*k}|^+ = \sum_{m=1}^{T-2} \max\{\beta_{*k}(t_m) - 2 \beta_{*k}(t_{m+1}) + \beta_{*k}(t_{m+2}), 0\}= 0$, i.e., $\beta_{*k}(t_m) - 2 \beta_{*k}(t_{m+1}) + \beta_{*k}(t_{m+2}) \leq 0$ for all $m\in[T-2]$. Therefore if, for example, we wanted to require concavity, we could use \eqref{eq:oracle_reframe} with the constraint $h^{(1)}(\mbbeta_k;M_k) = 0$ replaced with $|\mbD^{(2)}\mbbeta_{k}|^+ = 0$. To achieve nearly-concave estimates, we can relax this constraint, allowing $|\mbD^{(2)}\mbbeta_{k}|^+  \leq c$ for some positive constant $c$. Like the nearly-unimodal penalty, if $|\mbD^{(2)}\hat\mbbeta_{k}|^+ > 0$, then $\mbbeta_{k}$ is not concave. 

Based on these ideas, we propose the nearly-concave quantile distributed lag estimator 
\begin{equation}\label{eq:nearlyConcave}
\argmin_{\boldsymbol{\beta}, \mbgamma}~ \left[\frac{1}{n}\sum_{i=1}^n \rho_{\tau}\left\{ y_i- {\rm tr}(\mbX_i^\top \mbbeta) - \mbZ_i^\top\boldsymbol\gamma\right\} + \lambda_1 \sum_{k=1}^K  |\mbD^{(2)}\mbbeta_{k}|^+ + \lambda_2 \sum_{k=1}^K \|\mbD^{(2)}\mbbeta_{k}\|_2^2\right]
\end{equation}
where $\lambda_1 \geq 0$ and $\lambda_2 > 0$ are user-specified tuning parameters.  Compared to \eqref{eq:nearlyUnimodal}, \eqref{eq:nearlyConcave} does not require that we estimate the $M_k$ explicitly, and is the solution to a convex optimization problem. Due to this, \eqref{eq:nearlyConcave} is generally faster to compute than \eqref{eq:nearlyUnimodal}. 

\begin{figure}[t]
  \begin{center}
  \includegraphics[width=0.9\textwidth]{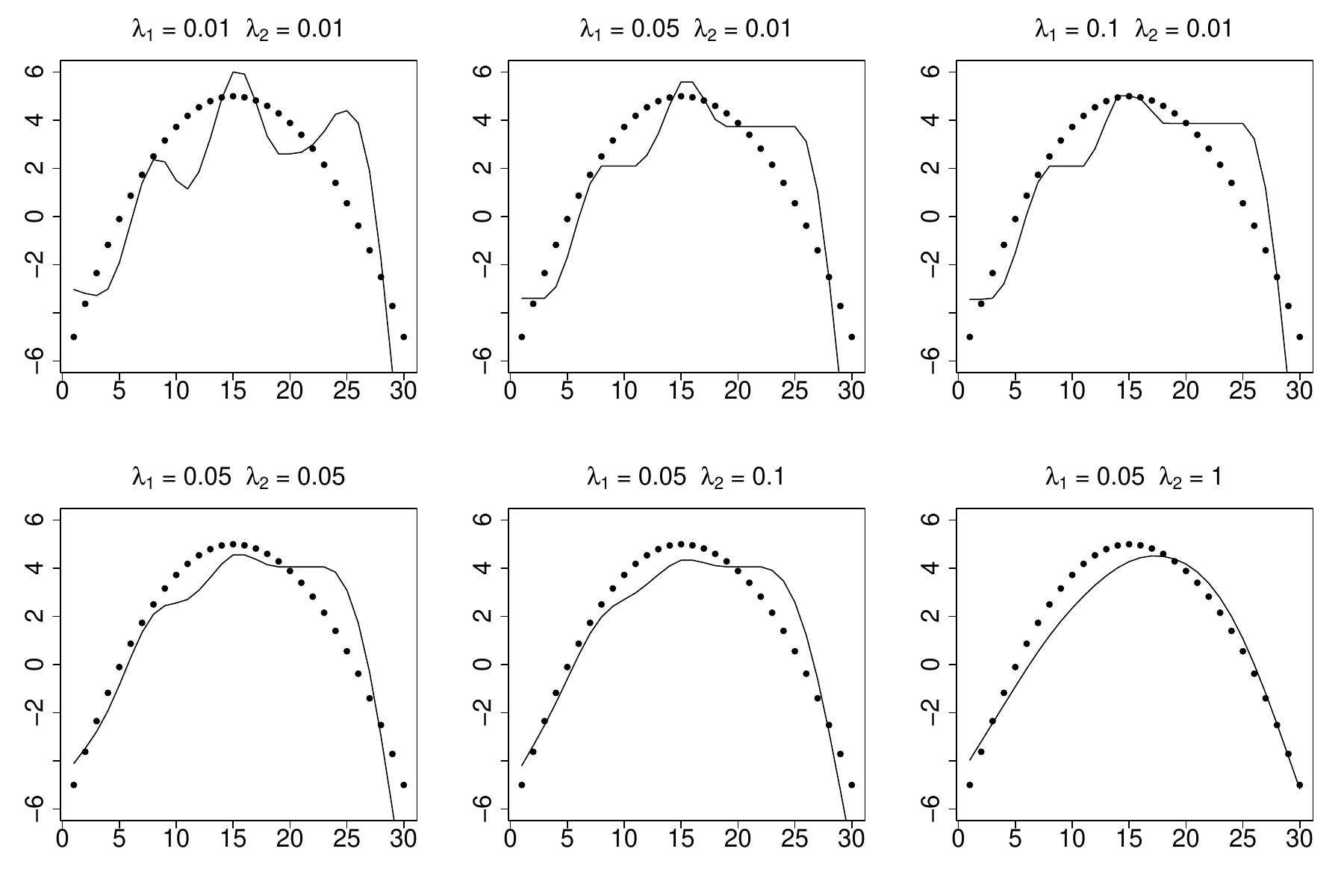}
  \end{center}
    \caption{Effects of the tuning parameters $\lambda_1$ and $\lambda_2$ on the nearly-unimodal quantile distributed lag estimator \eqref{eq:nearlyUnimodal}. The true $\mbbeta_*$ is shown in black dots, and  $\hat\mbbeta_*$ is represented by curves.}
  \label{fig::tuningPara1}
  \end{figure}

\begin{figure}[t]
    \begin{center}
    \includegraphics[width=0.9\textwidth]{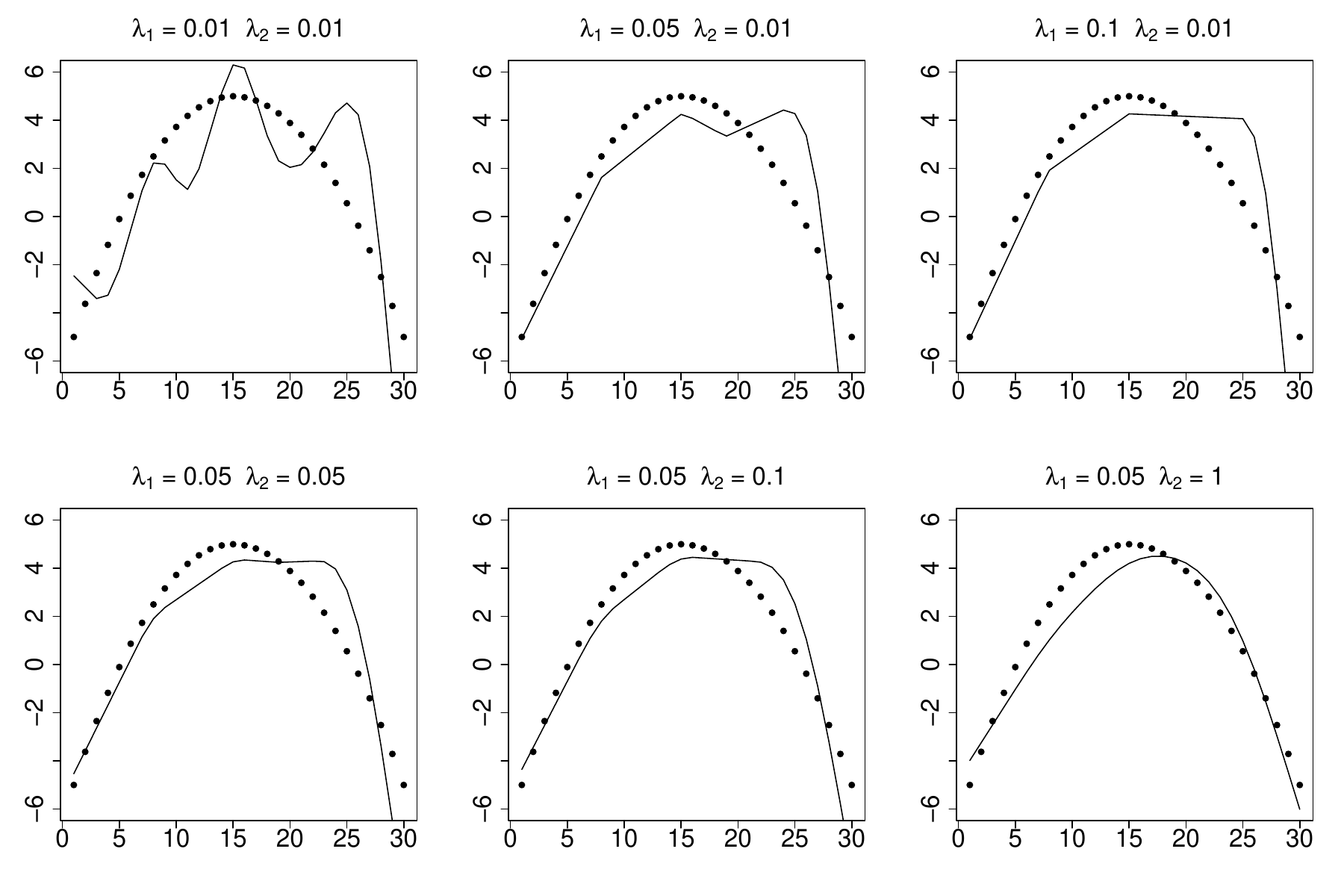}\end{center}
  \caption{Effects of the tuning parameters $\lambda_1$ and $\lambda_2$ on the nearly-concave quantile distributed lag estimator \eqref{eq:nearlyConcave}. The true $\mbbeta_*$ is shown in black dots, and $\hat\mbbeta_*$ is represented by curves.}
  \label{fig::tuningPara2}
\end{figure}

We illustrate the effects of the two tuning parameters, $\lambda_1$ and $\lambda_2$, on the nearly-unimodal and nearly-concave quantile distributed lag estimators in Figure \ref{fig::tuningPara1} and \ref{fig::tuningPara2}. Here we generated data from the quantile regression model with $K = 1$ and $T = 30.$ In these examples, we observe that with a fixed $\lambda_2$, increasing $\lambda_1$ enforces a stronger unimodal constraint on the nearly-unimodal estimator and a stronger concave constraint on the nearly-concave estimator. This is evident from the more pronounced unimodality as $\lambda_1$ increases. Conversely, with a fixed $\lambda_1$, increasing $\lambda_2$ results in smoother estimates across time for both the nearly-unimodal and nearly-concave estimators. This smoothness is achieved by reducing fluctuations in the estimated curves, demonstrating the role of $\lambda_2$ in controlling the smoothness of the solution. Together, tuning $\lambda_1$ and $\lambda_2$ allows for flexibility in shaping the estimated curves, balancing between the shape constraints and smoothness. By selecting $\lambda_1$ and $\lambda_2$ using cross-validation, we allow the best balance to be determined by the data. 

Note that if the time points of measurement $t_1, \dots, t_T$ differed across exposures, our method could still be applied straightforwardly. In this case, the $\mbD^{(v)}$ matrices for each $k \in [K]$ would need to be chosen in accordance with the number of time points. Moreover, if time points are not equally spaced, the rows of the corresponding $\mbD$ can be modified according to their spacing (e.g., by imposing a less harsh penalty on the difference between coefficients whose time points are farther apart).

\subsection{Related work}
Regularized quantile regression has been well-studied in the literature \citep[e.g., see Chapter 15 of ][ and references therein]{regression2017handbook}. For example, \citet{gu2018admm} proposed a variation of the  alternating direction method of multipliers (ADMM)  algorithm \citep{boyd2011distributed} for computing elastic-net penalized quantile regression estimators. To both alleviate the unhelpful bias often induced by convex penalties and handle the nondifferentiability of the check loss, \citet{tan2022high} developed a smooth approximation to the check loss to be used with folded-concave penalization. \citet{brantley2020baseline} studied quantile trend filtering, where they developed an efficient algorithm to compute 
$$ \argmin_{\mbb \in \mathbb{R}^T}\left\{  \frac{1}{2}\sum_{i=1}^T \rho_\tau\{u(t_i) - b(t_i)\} + \lambda_0 \|\mbD^{(1)}\mbb\|_1\right\},$$
which can provide a smooth approximation to the  $\tau$th quantile of the sequence $u(t_1), \dots, u(t_T)$ for appropriately selected tuning parameter $\lambda_0 > 0.$

In distributed lag generalized linear models, it is common to assume smoothness of the effects across time.  Smoothness is often achieved using penalized splines \citep{zanobetti2000generalized}, Gaussian processes \citep{warren2012spatial}, or utilizing ridge-type penalties on differences between coefficients corresponding to successive time points  \citep{obermeier2015flexible}. \citet{chen2018robust} consider multiple ways---some Bayesian---to impose smoothness. To the best of our knowledge, none of these methods impose shape constraints, nor do these methods easily extend to the quantile distributed lag case. 

From a methodological perspective, the work most closely related to our own is that on ``nearly-isotonic'' trend filtering \citep{tibshirani2011nearly}. \citet{tibshirani2011nearly} considered the problem of approximating a sequence $u(t_1), \dots, u(t_T)$ with a nearly-monotonically increasing sequence. Specifically, they proposed to approximate the sequence $u(t_1), \dots, u(t_T)$ using 
\begin{equation}\label{eq:nearly-iso}
\argmin_{\mbb \in \mathbb{R}^T}\left\{ \frac{1}{2}\sum_{i=1}^T \{u(t_i) - b(t_i)\}^2 + \lambda_0 |\mbD^{(1)}\mbb|^+\right\}.
\end{equation}
They also consider approximating the sequence with a nearly-convex function by replacing the penalty in \eqref{eq:nearly-iso}  with $\lambda_0 |\mbD^{(2)}\mbb|^-$. \citet{tibshirani2011nearly} proposed an efficient algorithm for computing the solution path to \eqref{eq:nearly-iso}, but did not consider application to quantile trend filtering, did not consider the case with predictors, nor did they consider estimating unimodal functions.

\section{Computation}
\subsection{Overview}
The optimization problems for the nearly-unimodal estimator \eqref{eq:nearlyUnimodal} and the nearly-concave estimator \eqref{eq:nearlyConcave} require different approaches due to their distinct sets of optimization variables. The optimization problem \eqref{eq:nearlyUnimodal} is nonconvex due to the discontinuity of the unimodal penalty term \(\overline{h}^{(1)}(\mbbeta_k; M_k)\) with respect to $M_k$. To tackle this nonconvex optimization problem, we propose using a blockwise coordinate descent algorithm. The key idea is to alternately update the two sets of parameters $(\mbbeta, \mbgamma)$ and $\{ M_k \}_{k=1}^K$ until convergence. Specifically, we use the following blockwise descent scheme. 

\noindent \rule{\textwidth}{0.4pt}\vspace{-2pt}
\noindent \textbf{Algorithm 1} Computing the nearly-unimodal quantile distributed lag estimator\vspace{-8pt}\\
\noindent \rule{\textwidth}{0.4pt}
\begin{enumerate}
  \setcounter{enumi}{-1}
  \vspace{-5pt}
    \item Initialize \(\mbbeta\) and \(\mbgamma\)
    \item Fix \(\mbbeta\) and \(\mbgamma\), and update \(\{M_k\}_{k=1}^K\)
    \item Fix \(\{M_k\}_{k=1}^K\), and update \(\mbbeta\) and \(\mbgamma\)
    \item Repeat Steps 1 and 2 until the algorithm converges\vspace{-12pt}
\end{enumerate}
\noindent  \rule{\textwidth}{0.4pt}

Fortunately, updating \(\{M_k\}_{k=1}^K\) in Step 1 is simple because each $M_k$ takes a value from the finite set of integers $[T]$. With \(\mbbeta\) and \(\mbgamma\) fixed at $\mbbeta^{(t)}$ and $\mbgamma^{(t)},$ the global minimizer of the objective function from \eqref{eq:nearlyUnimodal} with respect to \(M_k\), is given by
$$
M_k^{(t+1)} \in \argmin_{m \in [T]}~ \overline{h}^{(1)}(\mbbeta^{(t)}_k; m).$$
Since \(m\) only takes a finite number of values, we can evaluate \(\overline{h}^{(1)}(\mbbeta^{(t)}_k, m)\) for each possible value of \(m \in [T]\) and choose the one that yields the smallest value. In the (rare) case of ties, one can randomly choose an element from the set of minimizers. 

The optimization problem for Step 2 of Algorithm 1 and for computing \eqref{eq:nearlyConcave} have similar structures, so we propose an algorithm that can be used to solve both. Both problems can be written as
\begin{equation}
    \argmin_{\mbbeta, \mbgamma}\left[ \frac{1}{n}\sum_{i=1}^n\rho_{\tau}\{y_i-{\rm tr}(\mbX_i^\top \mbbeta) -\mbZ_i ^\top\mbgamma\} + \lambda_1  g(\mbbeta) + \lambda_2\sum_{k=1}^K \|\mbD^{(2)}\mbbeta_k\|^2\right],
    \label{admm}
\end{equation}
where $g(\mbbeta) = \sum_{k=1}^K \overline{h}^{(1)}(\mbbeta_k, M_k)$ for the nearly-unimodal estimator, and $g(\mbbeta) = \sum_{k=1}^K|\mbD^{(2)}\mbbeta_k|^+$ for the nearly-concave estimator. In the next subsection, we propose an ADMM algorithm to compute \eqref{admm} with either $g$. 

To simplify notation in the following subsection, let $\boldsymbol{Z} = (\mbZ_1, \dots, \mbZ_n)^\top \in \mathbb{R}^{n \times p}, \boldsymbol{y} = (y_1, \dots, y_n)^\top \in \mathbb{R}^n, \mbB = {\rm vec}(\mbbeta) \in \mathbb{R}^{TK}$, and ${\boldsymbol{X}} = ({\rm vec}(\mbX_1), \dots, {\rm vec}(\mbX_n))^\top \in \mathbb{R}^{n \times TK}$, where ${\rm vec}$ is the operator that stacks the columns of its matrix-valued argument. 

\subsection{Prox-linear ADMM algorithm for \eqref{admm}}
Solving \eqref{admm} is difficult because the quantile loss function and $g$ are nondifferentiable. The ADMM algorithm allows us to separate the quantile loss and penalties so that each can be dealt with separately through their so-called proximal operators. The proximal operator of a function $q$ is given by 
$$ {\rm \bf Prox}_q(\mbv) = \argmin_{\mbu} \left\{ \frac{1}{2}\|\mbu - \mbv\|_2^2 + q(\mbu)\right\}.$$
If $q$ is a closed proper convex function, its proximal operator is unique.  

To see how we can apply the ADMM algorithm, notice that the optimization problem in \eqref{admm} can be expressed as the constrained problem
\begin{equation}  \label{qADMM}
    \minim_{\boldsymbol{r}, \mbbeta, \mbgamma} ~ \left\{ \frac{1}{n}\sum_{i=1}^n\rho_{\tau}(\boldsymbol{r}_i) + \lambda_1 g(\mbbeta) +  \lambda_2\|\mbD\mbB\|_2^2 \right\}~~ {\rm subject~to} ~~ \mbr = \boldsymbol{y} - {\boldsymbol{X}}\mbB  - \boldsymbol{Z}\mbgamma,
    \end{equation}
    where $\mbD = \text{BlockDiag}(\{\mbD^{(2)}\}_{k=1}^K),$ 
so that $\sum_{k=1}^K\|\mbD^{(2)}\mbbeta_k\|_2^2=\|\mbD\mbB\|_2^2$. 
The ADMM algorithm solves \eqref{qADMM} by combining dual ascent with the method of multipliers. 
Let $f(\mbr)=\frac{1}{n}\sum_{i=1}^n\rho_{\tau}(\mbr_i)$, and $\rho > 0$ be a fixed step size. The augmented Lagrangian associated with \eqref{qADMM} is
$$
\begin{aligned}
  \mathcal L_\rho(\mbr, \mbbeta, \mbgamma, \mbu) = &f(\mbr) + \lambda_1g(\mbbeta) + \lambda_2\|\mbD\mbB\|^2_2 \\ 
  &-\langle \mbu, {\boldsymbol{X}}\mbB  + \mbZ\mbgamma+ \mbr - \mby\rangle + \frac{\rho}{2}\|{\boldsymbol{X}}\mbB  + \mbZ\mbgamma+ \mbr - \mby\|^2_2,
\end{aligned}
$$
where $\mbu \in \mathbb{R}^n$ is a dual variable. We use a modified version of the ADMM algorithm, sometimes called a ``prox-linear'' \citep{deng2016global} or ``proximal'' ADMM algorithm \citep{gu2018admm}. The prox-linear ADMM algorithm is especially useful when subproblems of the standard ADMM algorithm require minimizing a penalized least squares criterion.  This algorithm has $(t+1)$th iterates
\begin{align}
  \mbbeta^{(t+1)}, \mbgamma^{(t+1)} &= \argmin_{\mbbeta, \mbgamma} \bigg\{ \mathcal L_\rho(\mbr^{(t)}, \mbbeta,\mbgamma, \mbu^{(t)}) + \frac{1}{2}\|\mbB-\mbB^{(t)}\|^2_{\mbS}  \bigg\}\label{bg-update} \\
  \mbr^{(t+1)} &= \argmin_{\mbr} \left\{ \mathcal L_\rho(\mbr, \mbbeta^{(t+1)},\mbgamma^{(t+1)}, \mbu^{(t)})\right\} \label{r-update} \\
  \mbu^{(t+1)} &= \mbu^{(t)} - \rho({\boldsymbol{X}}\mbB^{(t+1)}  + \mbZ\mbgamma^{(t+1)}+ \mbr^{(t+1)} - \mby) \label{u-update}
\end{align}
where $\mbS$ is a positive semidefinite matrix to be defined shortly, and $\|\mbv\|_{\mbS}^2 = \langle \mbv, \mbS \mbv\rangle$ is the seminorm induced by the semi-inner product defined via $\mbS$. The standard ADMM algorithm \citep[Chapter 3][]{boyd2011distributed} replaces \eqref{bg-update} with $\argmin_{\mbbeta, \mbgamma} \mathcal L_\rho(\mbr^{(t)}, \mbbeta,\mbgamma, \mbu^{(t)})$. Because of this, the iterate defined in \eqref{bg-update} can be thought of as an approximation to the corresponding update from the vanilla ADMM algorithm.  By incorporating the quadratic term $\|\mbB-\mbB^{(t)}\|^2_{\mbS}$, the update for $\mbbeta$ in \eqref{bg-update} can be transformed into solving two independent nearly-isotonic regression problems for the nearly-unimodal estimator and solving a single nearly-concave regression problem for the nearly-concave estimator.
Steps \eqref{r-update} and \eqref{u-update} follow the vanilla ADMM algorithm.

To tackle \eqref{bg-update}, notice that after a bit of algebra, we can write 
$$
\begin{aligned}
  \mathcal L_\rho(\mbr^{(t)}, \mbbeta,\mbgamma, \mbu^{(t)})
  &= \lambda_1 g(\mbbeta) + \frac{\rho}{2}\|\widetilde{{\boldsymbol{X}}}\mbB+\widetilde\mbZ\mbgamma + \widetilde\mbr^{(t)}-\widetilde\mby -  \rho^{-1} \widetilde\mbu^{(t)}\|^2_2+ f(\mbr^{(t)})
\end{aligned}
$$
where $\widetilde{\boldsymbol{X}}=({\boldsymbol{X}}^\top, \sqrt{\frac{2\lambda_2}{\rho}}\mbD^\top)^\top\in\mathbb{R}^{(n+KT-2K)\times KT}, \widetilde\mbZ=(\mbZ^\top, \mathbf 0)^\top\in\mathbb{R}^{(n + KT - 2K)\times p},\widetilde{\mbr}^{(t)}= ({\mbr^{(t)}}^\top, \mathbf 0^\top)^\top\in\mathbb{R}^{n+KT-2K}, \widetilde\mby= ({\mby}^\top, \mathbf 0^\top)^\top\in\mathbb{R}^{n+KT-2K},$ $\widetilde\mbu^{(t)}= ({\mbu^{(t)}}^\top, \mathbf 0^\top)^\top\in\mathbb{R}^{n+KT-2K}$, and $f(\mbr^{(t)})$ is constant with respect to $(\mbbeta, \mbgamma)$.
Then for fixed $\mbbeta$, the minimizer of $\mathcal L_\rho(\mbr^{(t)}, \mbbeta,\mbgamma, \mbu^{(t)})$ with respect to $\mbgamma$ is given by
\begin{equation}
  \hat\mbgamma(\mbbeta) = (\widetilde\mbZ^\top\widetilde\mbZ)^{-1}\widetilde\mbZ^\top(\widetilde\mby - \widetilde\mbr^{(t)} + \rho^{-1}\widetilde\mbu^{(t)} - \widetilde{\boldsymbol{X}}\mbB).
  \label{gamma}
\end{equation}
Plugging $\hat\mbgamma(\mbbeta)$ back into $\mathcal L_\rho(\mbr^{(t)}, \mbbeta,\mbgamma, \mbu^{(t)})$, we have
$$\mathcal L_\rho(\mbr^{(t)}, \mbbeta,\hat\mbgamma(\mbbeta), \mbu^{(t)})=\lambda_1 g(\mbbeta)+\frac{\rho}{2}\|\widetilde{\boldsymbol{X}}_{\widetilde\mbZ}\mbB -\bar\mbt^{(t)}\|^2_2+ f(\mbr^{(t)})$$
where $\widetilde{\boldsymbol{X}}_{\widetilde\mbZ}=(I_{n+KT-2K}-P_{\widetilde\mbZ})\widetilde{\boldsymbol{X}}$, $\bar\mbt^{(t)}=(I_{n+KT-2K}-P_{\widetilde\mbZ})(\widetilde\mby - \widetilde\mbr^{(t)}+\rho^{-1}\widetilde\mbu^{(t)})$ and $P_{\widetilde\mbZ}=\widetilde\mbZ(\widetilde\mbZ^\top\widetilde\mbZ)^{-1}\widetilde\mbZ^\top$.

Now, we are ready to derive the update for $\mbbeta$. Defining $\mbS=\rho(\eta I_{KT}-\widetilde{\boldsymbol{X}}_{\widetilde\mbZ}^\top\widetilde{\boldsymbol{X}}_{\widetilde\mbZ}), ~ {\rm with}~ \eta \in \mathbb{R}$ fixed at a value greater than or equal to the largest eigenvalue of $\widetilde{\boldsymbol{X}}_{\widetilde\mbZ}^\top\widetilde{\boldsymbol{X}}_{\widetilde\mbZ}$, we have
\begin{align*}
  \mbbeta^{(t+1)}
  &= \argmin_{\mbbeta\in\mathbb R^{K \times T}} \left\{ \lambda_1 g(\mbbeta) + \frac{\rho}{2}\|\widetilde{\boldsymbol{X}}_{\widetilde\mbZ}\mbB -\bar\mbt^{(t)}\|^2_2  + \frac{1}{2}\|\mbB-\mbB^{(t)}\|^2_{\mbS}\right\}\\
  &= \argmin_{\mbbeta\in\mathbb R^{K \times T}} \sum_{k=1}^K\left\{\frac{\lambda_1}{\rho\eta}g(\mbbeta_k) + \frac{1}{2}\|\mbbeta_k - \mbs^{(t)}_k\|^2_2\right\}
\end{align*}
where $({\mbs^{(t)}_1}^\top, \dots, {\mbs^{(t)}_K}^\top)^\top = \mbB^{(t)} + \eta^{-1}\widetilde{\boldsymbol{X}}_{\widetilde\mbZ}^\top(\bar\mbt^{(t)}-\widetilde{\boldsymbol{X}}_{\widetilde\mbZ}\mbB^{(t)})$ with $g(\mbbeta_k) = \overline{h}^{(1)}(\mbbeta_k, M_k)$ for the nearly-unimodal estimator and $g(\mbbeta_k) = |\mbD^{(2)}\mbbeta_k|^+$ for the nearly-concave estimator. Restated more simply, we see that for each $\mbbeta_k$, 
$$\mbbeta^{(t+1)}_k = {\rm \bf Prox}_{\frac{\lambda_1}{\rho\eta} g}(\mbs_k^{(t)}).$$

From the derivation above, one can see that our particular choice of $\mbS$ led to cancellations with other quadratic terms so that the update for $\mbbeta$ reduces to the proximal operator of the (scaled) function $g$. Furthermore, by taking $\eta$ greater than or equal to the largest eigenvalue of $\widetilde{\boldsymbol{X}}_{\widetilde\mbZ}^\top\widetilde{\boldsymbol{X}}_{\widetilde\mbZ}$, we are ensured that 
$\mathcal L_\rho(\mbr^{(t)}, \mbbeta^{(t+1)},\hat\mbgamma(\mbbeta^{(t+1)}), \mbu^{(t)}) \leq \mathcal L_\rho(\mbr^{(t)}, \mbbeta^{(t)},\hat\mbgamma(\mbbeta^{(t)}), \mbu^{(t)})$
by the majorize-minimize principle \citep{lange2016mm}, which is essentially for the convergence of our algorithm.

For the nearly-unimodal estimator, the proximal operator of $(\lambda_1/\rho\eta)g$ at $\mbs_k^{(t)} = (s_{k1}^{(t)}, \dots, s_{kT}^{(t)})^\top$ can be expressed as
\begin{align*}
\mbbeta^{(t+1)}_k &= \argmin_{\mbb \in\mathbb R^T} \left[\frac{1}{2}\sum_{i=1}^{M_k}\{s^{(t)}_{ki}- b(t_i)\}^2 + \frac{\lambda_1}{\rho\eta}\sum_{i=1}^{M_k-1}\max\{b(t_i) - b(t_{i+1}), 0\}\right.\\
& ~~~~~~~~~~~~~\quad\quad\quad \left. + \frac{1}{2}\sum_{i=M_k+1}^{T}\{s^{(t)}_{ki}-b(t_{i})\}^2 + \frac{\lambda_1}{\rho\eta}\sum_{i=M_k+1}^{T-1}\max\{b(t_{i+1}) - b(t_{i}), 0\}\right]
\end{align*}
for $k \in [K]$, where $\mbb = (b(t_1), \dots, b(t_T))^\top$. Thus, the update for $\mbbeta^{(t+1)}_k$ can be obtained by fitting two separate nearly-isotonic regression models \eqref{eq:nearly-iso} with tuning parameter $\lambda_1/(\rho\eta)$, one for $\mbbeta_{k,1:M_k}$ based on $(s^{(t)}_{k1}, \dots, s^{(t)}_{k M_k})^\top$ and the other for $\mbbeta_{k,(M_k+1):T}$ based on $(s^{(t)}_{kM_{k+1}}, \dots, s^{(t)}_{k T})^\top$. For these, we use the algorithm from \citet{tibshirani2011nearly}, which is very efficient for \eqref{eq:nearly-iso}. 

For the nearly-concave estimator, the update simplifies to
\begin{equation}
  \mbbeta^{(t+1)}_k = \argmin_{\mbb \in\mathbb R^T} \left[\frac{1}{2}\sum_{i=1}^{T}\{s^{(t)}_{ki}- b(t_i)\}^2 + \frac{\lambda_1}{\rho\eta}\sum_{i=1}^{m-1}\max\{b(t_i) - 2b(t_{i+1}) + b(t_{i+2}), 0\} \right]
  \label{admm2}
\end{equation}
for $k \in[K]$, which can be computed efficiently using an ADMM sub-algorithm. We provide this sub-algorithm in the Supplementary Materials. After we compute $\mbbeta^{(t+1)}$, we can update $\mbgamma^{(t+1)}$ using \eqref{gamma}.

It can be shown that $\mbr^{(t+1)}$ has a closed-form solution \citep[e.g., see ][]{gu2018admm}. In fact, the update of $\mbr^{(t+1)} = (r_1^{(t+1)}, \dots, r_n^{(t+1)})^\top$ can be carried out component-wise in parallel. For $i \in [n]$, one can show that 
$$
\begin{aligned}
  r^{(t+1)}_i 
  &= {\rm \bf Prox}_{\rho_{\tau}/n\rho}\big\{y_i- {\rm tr}(\mbX_i^\top\mbbeta^{(t+1)}) - \mbZ_i^\top\mbgamma^{(t+1)}+\rho^{-1}u^{(t)}_i\big\}
\end{aligned}
$$
where 
$$
{\rm \bf Prox}_{\rho_{\tau}/\alpha}(\xi)=\left\{
  \begin{aligned}
    &\xi - \frac{\tau}{\alpha}, & &:~ \xi > \tau/\alpha \\
    &\xi - \frac{\tau-1}{\alpha} & &:~ \xi < (\tau-1)/\alpha   \\
    &0, & &:{\rm otherwise}
  \end{aligned}
\right.
$$

The ADMM algorithm updates are repeated until a stopping criterion is reached. We use the criterion based on that outlined in Section 3.3.1 of \cite{boyd2011distributed}. Specifically, the algorithm terminates when both the relative primal and dual residuals fall below a certain tolerance threshold, that is,
$$
\begin{aligned}
  \|\mbX\mbB^{(t)} + \mbZ\mbgamma^{(t)}+\mbr^{(t)}-\mby\|_2 &\le \sqrt{n}\epsilon_1 + \epsilon_2\max\{\|\mbX\mbB^{(t)} + \mbZ\mbgamma^{(t)}\|_2, \|\mbr^{(t)}\|_2, \|\mby\|_2\}, \\
  \rho\|\mbW^\top(\mbr^{(t)}-\mbr^{(t-1)})\|_2&\le \sqrt{T+p}\epsilon_1 + \epsilon_2\|\mbW^\top\mbu^{(t)}\|_2
\end{aligned}
$$
where $\mbW = (\mbX, \mbZ)\in\mathbb{R}^{n\times(p+KT)}$. We set $\epsilon_1=10^{-4}$ and $\epsilon_2=10^{-4}$. The algorithm for solving \eqref{admm} is summarized below. 
\setcounter{algorithm}{1}
  \begin{algorithm}[H]
  \begin{algorithmic}[1]
  \STATE Initialize the algorithm with ($\mbbeta^{(0)}, \mbgamma^{(0)}, \mbr^{(0)}, \mbu^{(0)}$).
  \FOR{$t = 1, 2, \dots$ until convergence}
  \STATE Update $\mbbeta^{(t+1)} \leftarrow \argmin_{\mbbeta\in\mathbb R^{K \times T}} \{\lambda_1 g(\mbbeta) + \frac{\rho\eta}{2}\|\mbB-  \mbB^{(t)} - \eta^{-1}\widetilde\mbX_{\widetilde\mbZ}^\top(\bar\mbt^{(t)}-\widetilde\mbX_{\widetilde\mbZ}\mbB^{(t)})\|^2_2\}$
  \STATE Update $\mbgamma^{(t+1)} \leftarrow (\widetilde\mbZ^\top\widetilde\mbZ)^{-1}\widetilde\mbZ^\top(\widetilde\mby - \widetilde\mbr^{(t)} + \rho^{-1}\widetilde\mbu^{(t)} - \widetilde\mbX\mbB^{(t+1)})$
  \STATE Update $r_i^{(t+1)} \leftarrow {\rm \bf Prox}_{\rho_{\tau}/n\rho}\{y_i-\mbX_i^\top\mbbeta^{(t+1)} - \mbZ_i^\top\mbgamma^{(t+1)}+\rho^{-1}u^{(t)}_i\}$ for each $i \in [n]$
  \STATE Update $\mbu^{(t+1)} \leftarrow \mbu^{(t)} - \rho({\boldsymbol{X}}\mbB^{(t+1)}  + \mbZ\mbgamma^{(t+1)}+ \mbr^{(t+1)} - \mby)$
  \ENDFOR
  \end{algorithmic}
  \caption{Prox-linear ADMM algorithm for solving penalized quantile regression \eqref{admm}}
  \label{alg:seq}
  \end{algorithm}

We can show that the iterates of our ADMM algorithm converge to a global minimizer. In particular,we can apply the results of \citet{gu2018admm}, which are stronger than standard convergence results for the ADMM algorithm. 
\begin{remark}
  If we let $\mbtheta=(\mbB^\top, \mbgamma^\top)^\top$ and define 
  $\widetilde\mbS=\begin{pmatrix}
    \mbS & \mathbf 0 \\
    \mathbf 0 & \mathbf 0
  \end{pmatrix}$
  where $\mbS$ is defined as above, we can rewrite \eqref{bg-update} as 
  $
    \mbtheta^{(t+1)} = \argmin_{\mbtheta} \mathcal L_\rho(\mbr^{(t)}, \mbtheta, \mbu^{(t)}) + \frac{1}{2}\|\mbtheta-\mbtheta^{(t)}\|^2_{\widetilde\mbS}.$
  Then, by identical arguments as in the proof Theorem 1 from \cite{gu2018admm}, we can show that the sequence $\{(\mbtheta^{(t)}, \mbr^{(t)}), t=0,1,2,\dots\}$ generated by the prox-linear ADMM algorithm converges to an optimal solution $\{\mbtheta^\star, \mbr^\star\}$ of \eqref{qADMM}, and $\{\mbu^{(t)}, t=0,1,2,\dots\}$ converges to an optimal solution $\mbu^\star$ to the dual problem of \eqref{qADMM}.
\end{remark}

Software implementing both estimators, along with code for reproducing simulation study results, can be downloaded from \url{https://github.com/yjin07/smoothQDLM}.

\section{Inference using the wild bootstrap}
In order to perform approximate inference with our estimator, we propose to use a wild bootstrap procedure. 
Traditional bootstrap methods, such as the residual bootstrap or paired bootstrap, can fail to account for the  heteroscedasticity inherent in quantile regression models. These methods may lead to biased variance estimates and invalid inference.
In contrast, the wild bootstrap, introduced by \citet{wu1986jackknife} and \citet{liu1988bootstrap}, and later developed for quantile regression by \citet{feng2011wild}, offers a more robust alternative. \citet{wang2018wild} showed that for an adaptive $L_1$-penalized quantile regression, the wild bootstrap of \citet{feng2011wild} can be asymptotically valid for approximating the distribution of their penalized quantile regression estimator. We adopt the same approach as \citet{wang2018wild} in this work. 

To implement the wild bootstrap for our penalized quantile regression model---focusing on \eqref{eq:nearlyUnimodal} for concreteness---we use the following procedure.
\begin{enumerate}
  \item \textbf{Compute $\hat\mbbeta$ and $\hat\mbgamma$.} Compute estimates $\hat\mbbeta$ and $\hat\mbgamma$ using the penalized quantile regression model \eqref{eq:nearlyUnimodal} with tuning parameters chosen by cross-validation. 
  \item \textbf{Calculate residuals.} Compute the residuals $\hat \mbe_i=y_i-{\rm tr}(\mbX_i^\top\hat\mbbeta)-\mbZ_i^\top\hat\mbgamma$ for $i \in [n].$
  \item \textbf{Generate bootstrap samples.} For each $i \in [n]$ independently,
  \begin{enumerate}
    \item Generate weight $w_i$ as a realization of $W_i$, where ${\rm Pr}\{W_i = 2(1-\tau)\} = (1-\tau)$ and ${\rm Pr}(W_i = -2\tau) = \tau$.
    \item Create the bootstrap sample $y_i^\star= {\rm tr}(\mbX_i^\top\hat\mbbeta) +\mbZ_i^\top\hat\mbgamma+w_i\hat \mbe_i$
  \end{enumerate}
  \item \textbf{Refit the model to the bootstrap samples.} Using the bootstrap dataset $\{(\mby^\star_i, \mbX_i, \mbZ_i)\}_{i=1}^n$, compute bootstrap estimates $\hat\mbbeta^{\rm boot}$ and $\hat\mbgamma^{\rm boot}$ using \eqref{eq:nearlyUnimodal} with tuning parameters chosen by cross-validation. 
  \item \textbf{Repeat $B$ times.} Repeat Steps 3 and 4 $B$ times independently to obtain $B$ bootstrap estimates $\hat\mbbeta^{\rm boot}$ and $\hat\mbgamma^{\rm boot}$.
\item \textbf{Construct intervals.} 
  Based on the $B$ bootstrap estimates,  construct $(1-\alpha)$100\% confidence intervals for $\mbbeta_*$ and $\mbgamma_*$ according to the bootstrap empirical distribution of $\hat\mbbeta$ and $\hat\mbgamma$. 
\end{enumerate}
There are other distributions that can be used for sampling the weights to provide asymptotically valid inference: see \cite{feng2011wild} and \cite{wang2018wild} for more details.  

In practice, performing cross-validation for every bootstrap estimate can be too computationally burdensome to be used in practice. Instead, we approximate the bootstrap distribution by, in Step 4, estimating $\mbbeta_*$ and $\mbgamma_*$ on the bootstrap dataset using the tuning parameter pair selected in Step 1. This allows us to solve \eqref{eq:nearlyUnimodal} for only a single pair of tuning parameters, as opposed to computing the entire solution path multiple times in cross-validation. In general, we found that this had a minimal effect on our estimate of the distribution of $\hat\mbbeta$ and $\hat\mbgamma$.

\section{Simulation studies}
\subsection{Data generating models}
In this section, we conduct simulation studies to evaluate the performance of the proposed methods. We compare our estimators to competitors under a variety of data generating models. The model for the simulated data is
$$
  Y_i = {\rm tr}(\mbX_{i}^\top\mbbeta_{*})  + \mbZ_i^\top\mbgamma_* + \sigma\epsilon_i, \quad i \in [n],
$$
where each row of $\mbX_i$ is a realization of ${\rm N}_T(0, \mbSigma_X)$, $\mbZ_i\sim {\rm N}_p(0, \boldsymbol{I}_p)$. We set $\mbSigma_X= 0.8^{|j-k|}$ for all $(j,k)\in[T]\times[T]$ to create a high degree of temporal correlation in the exposures, which is expected in practice. The values of $T$, $p$ and $K$ are fixed at $30, 5,$ and $6$, respectively.

Two error distributions for the $\epsilon_i$ are used: (1) a standard normal distribution and (2) a $t$-distribution with 4 degrees of freedom. We vary $\sigma$ to control the signal-to-noise ratio (SNR). Three settings for generating the $\mbbeta_{*k}$ are considered.

\begin{itemize}
  \item[$\bullet$] \textbf{Model A}. For \(k \in [K]\), we generate \(\mbbeta_{*k}\) such that \(\beta_{*k}(t_1) \le \beta_{*k}(t_2) \le \cdots \le \beta_{*k}(t_{M_k}) \ge \cdots \ge \beta_{*k}(t_T)\). Specifically, we set \(\beta_{*k}(t_{M_k}) = 5\), \(\beta_{*k}(t_{1}) = \beta_{*k}(t_T) = -5\). The differences \(|\beta_{*k}(t_i) - \beta_{*k}(t_{i+1})| = c_i\) for \(i \in [T-1]\) are generated as follows: First, we generate random values for the differences between consecutive \(\beta\) values on both the increasing and decreasing segments of the sequence. These differences are scaled so that their cumulative sum equals \(10\) (the total range from \(-5\) to \(5\) and back to \(-5\)). The differences \(c_i\) are fixed and remain the same for every replication to ensure consistency. We set \((M_1, M_2, M_3, M_4, M_5, M_6) = (12, 15, 18, 17, 15, 13)\).

  \item[$\bullet$] \textbf{Model B}. First, we generate \(\widetilde\mbbeta_*\) according to Model A. Then, we modify the generated coefficients to introduce sparsity by setting \(\beta_{*k}(t_j) = \widetilde\beta_{*k}(t_j) \mathbf{1}(|\widetilde\beta_{*k}(t_j)| > 2.5)\) for \(k \in [K]\) and \(j \in [T]\).

  \item[$\bullet$] \textbf{Model C}. For $k \in [K]$, we generate $\mbbeta_{*k}$ such that the coefficients follow a parabolic shape. Specifically, we first generate indices for the sequence and compute parabolic values for each part. We divide the time points into two segments: $t_1$ to $t_{M_k}$ and $t_{M_k}$ to $t_T$. For the first segment ($t_1$ to $t_{M_k}$), we generate equally spaced values $x$ from $-1$ to $0$. For the second segment ($t_{M_k}$ to $t_T$), we generate equally spaced values $x$ from $0$ to $1$. Next, we compute the parabolic values for each part: For the first segment, the values are given by $y = 5x^2$. For the second segment, the values are given by $y = 5x^2$. We then combine the two segments to form a parabolic sequence that peaks at $t_{M_k}$. The sequence is adjusted so that the maximum value is 5 at $t_{M_k}$ and the minimum value is -5 at $t_1$ and $t_T$. This ensures a smooth and parabolic shape peaked at $M_k$. We set $(M_1, M_2, M_3, M_4, M_5, M_6) = (12, 15, 18, 17, 15, 13)$.
\end{itemize}

All three models generate $\mbbeta_{*k}$ coefficients that are unimodal in shape. The primary distinction lies in the structure of the sequences: while $\mbbeta_{*k}$ from Model A is strictly increasing and then strictly decreasing, $\mbbeta_{*k}$ from Model B incorporates zeros, introducing a sparsity aspect into the coefficients. Model C is unique in that it generates the coefficients $\mbbeta_{*k}$ so that they are not only unimodal, but also strictly concave in shape. Throughout the simulation study, we consider $n\in\{500, 750, 1000\}$ and ${\rm SNR}\in\{0.1, 0.25, 0.5\}$. The quantile $\tau$ is set to be $0.25$. We did not observe a substantial difference in relative performances between methods when using other quantiles, so additional results are omitted. 

\subsection{Competing methods and performance metric}
We compare our methods, \texttt{Uni} and \texttt{Concave}---\eqref{eq:nearlyUnimodal} and \eqref{eq:nearlyConcave}, respectively---to several competitors. The first competitor is the ridge-penalized quantile regression estimator, which we call (\texttt{Ridge}). 
The second competitor is the elastic-net penalized quantile regression estimator, (\texttt{EN}), defined as 
\begin{equation}
  \begin{aligned}
    \argmin_{\mbbeta, \mbgamma}\left[\frac{1}{n}\sum_{i=1}^n\rho_{\tau}\{\mby_i- {\rm tr}(\mbX_{i}^\top\mbbeta)  + \mbZ_i^\top\mbgamma\}  + \lambda\sum_{k=1}^K\left(\frac{1-\alpha}{2}\|\mbbeta_k\|^2_2 + \alpha\|\mbbeta_k\|_1\right)\right].
  \end{aligned}
  \label{en}
\end{equation}
The estimator \texttt{Ridge} is a  a special case of \texttt{EN} with $\alpha = 0$. 
The third competitor imposes smoothness conditions on the distributed lag curves across time.  
It assumes a functional form for the distributed lag curves $\mbbeta_{*k}=(\beta_{k}(t_1),\beta_{k}(t_2),\dots,\beta_{k}(t_T))^\top$ for $k \in [K]$. 
That is,
$$
\beta_{k}(t) = \sum_{j=1}^{l_k}f_{kj}(t)\xi_{kj} = \boldsymbol{f}^\top_k(t)\boldsymbol{\xi}_k
$$
wherein $f_{kj}(t)$ denotes basis functions, which must be determined in advance. We adopted the approach used in \cite{wilson2017bayesian} to construct data-driven basis functions for distributed lags as the eigenvectors of the covariance matrix the $n\times T$ matrix of exposure $k$ measured over time. To obtain a smooth orthonormal basis, we use fast
covariance estimation (FACE) proposed by \cite{xiao2016fast} to obtain the eigenfunctions of a
smoothed covariance matrix, as implemented in the R package refund \citep{Goldsmith2023}. The selection criterion for the number of basis functions, $l_k$, is determined by identifying the minimum integer required such that the cumulative variance accounted for by the initial $l_k$ eigenfunctions reaches or surpasses 90\%.  We denote this method as \texttt{FPCA}, and the coefficients are estimated by
\begin{equation}
  \argmin_{\mbgamma, \{\xi_k\}_{k=1}^K} \left[\sum_{i=1}^n\rho_{\tau}\{y_i-\sum_{k=1}^{K}\sum_{t=1}^T\boldsymbol{f}^\top_k(t)\boldsymbol{\xi}_k x_{ikt}  + \mbz_i^\top\mbgamma\}\right],
\end{equation}
where $x_{ikt}$ is the $(k,t)$th entry of the $\mbX_i.$
The fourth competitor is the same as \texttt{FPCA} but with a ridge penalty on the coefficients $\xi_k$'s. This method is referenced as \texttt{FPCA-R}. 
A separate validation set of size $n$ is used to select the tuning parameters for \texttt{Uni}, \texttt{Ridge}, \texttt{EN} and \texttt{FPCA-R}. For each setting, we generate 50 independent replications of the data and report the average performance of the estimators. The performance is measured by the estimation error of the coefficients, which is defined as $\|\boldsymbol\beta^* - \hat{\boldsymbol{\beta}}\|_2$.

\begin{figure}[t]
  \centering
  \begin{minipage}[t]{\textwidth}
  \centering
  \includegraphics[width=0.9\textwidth]{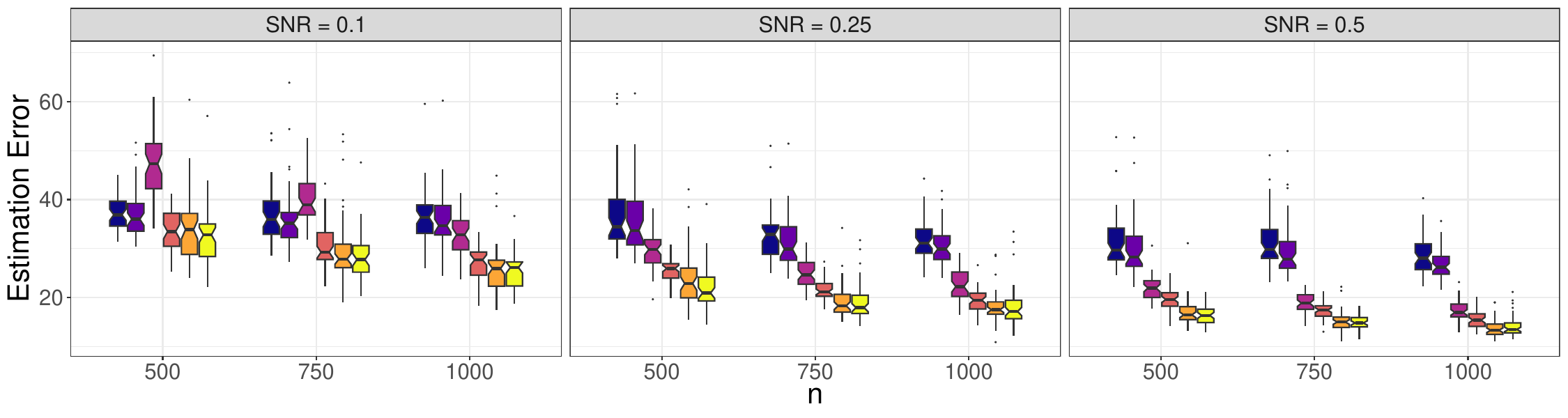}\\
  \end{minipage}
  \begin{minipage}[t]{\textwidth}
    \centering
    \includegraphics[width=0.9\textwidth]{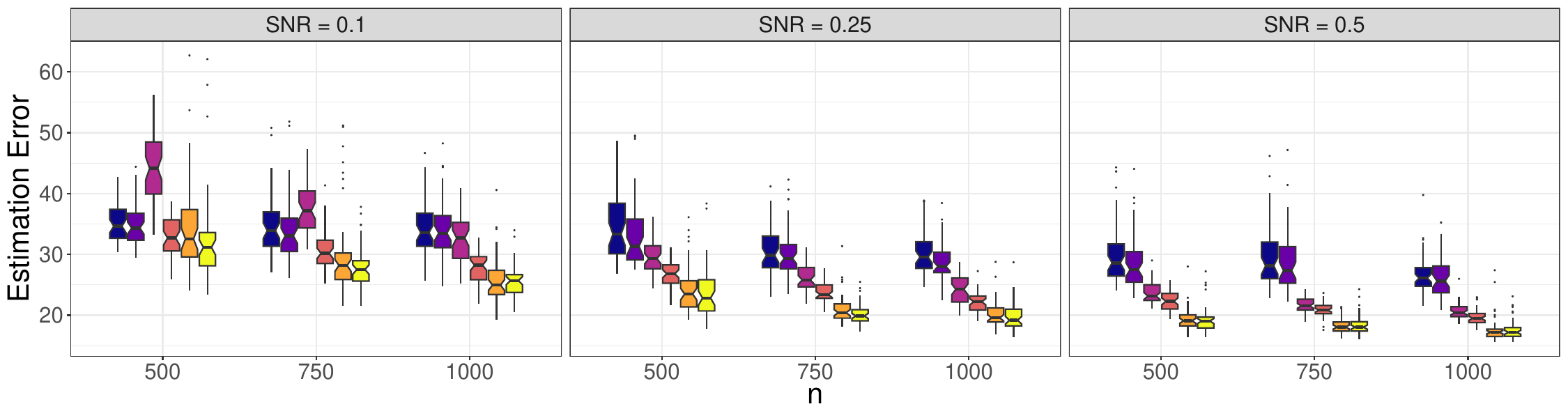}\\
    \end{minipage}
    \begin{minipage}[t]{\textwidth}
      \centering
      \includegraphics[width=0.9\textwidth]{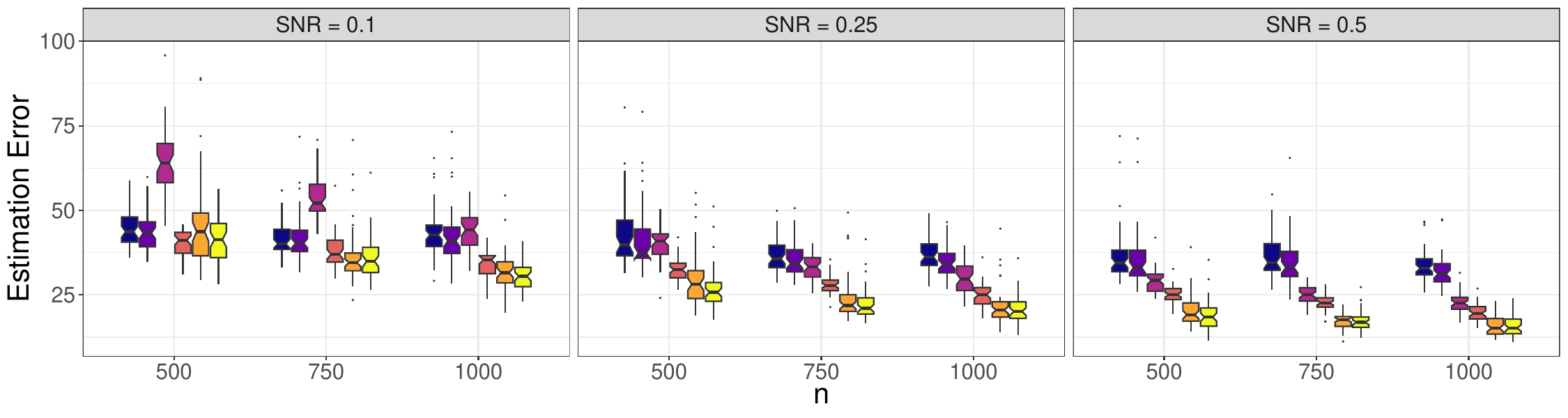}\\
      \end{minipage}
  \centering
  \includegraphics[width=6cm]{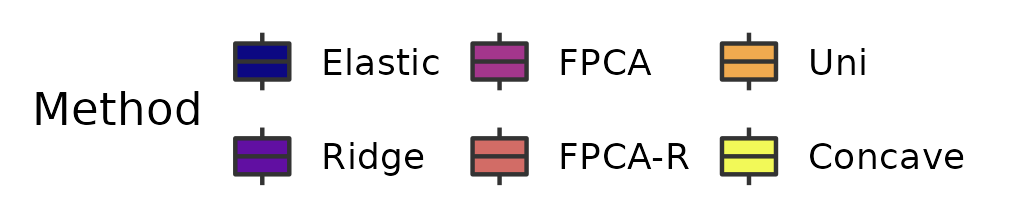}
  \caption{Estimation errors over 50 independent replications under \textbf{Model A} (first row), \textbf{Model B} (second row) and \textbf{Model C} (third row) with normal errors.}
  \label{fig::new}
\end{figure}

\begin{figure}[t]
  \centering
  \begin{minipage}[t]{\textwidth}
  \centering
  \includegraphics[width=0.9\textwidth]{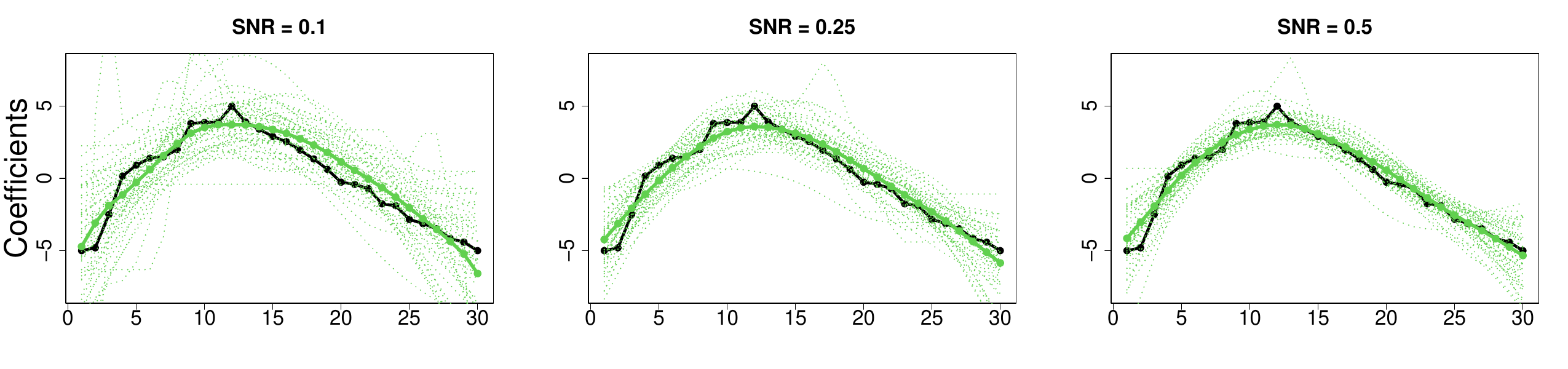}
  \end{minipage}
  \begin{minipage}[t]{\textwidth}
    \centering
    \includegraphics[width=0.9\textwidth]{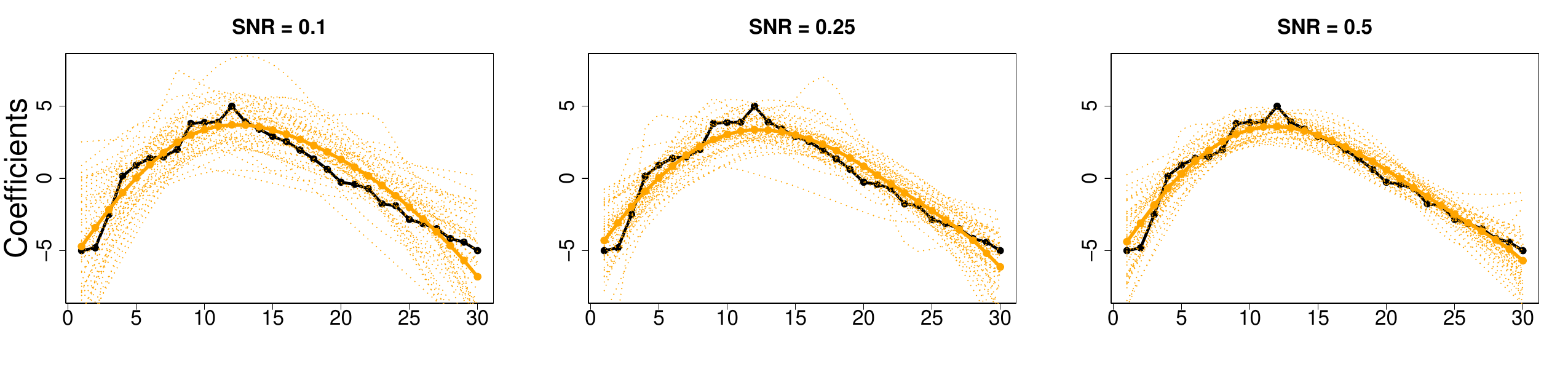}
    \end{minipage}
  \centering
  \begin{minipage}[t]{\textwidth}
  \centering
  \includegraphics[width=0.9\textwidth]{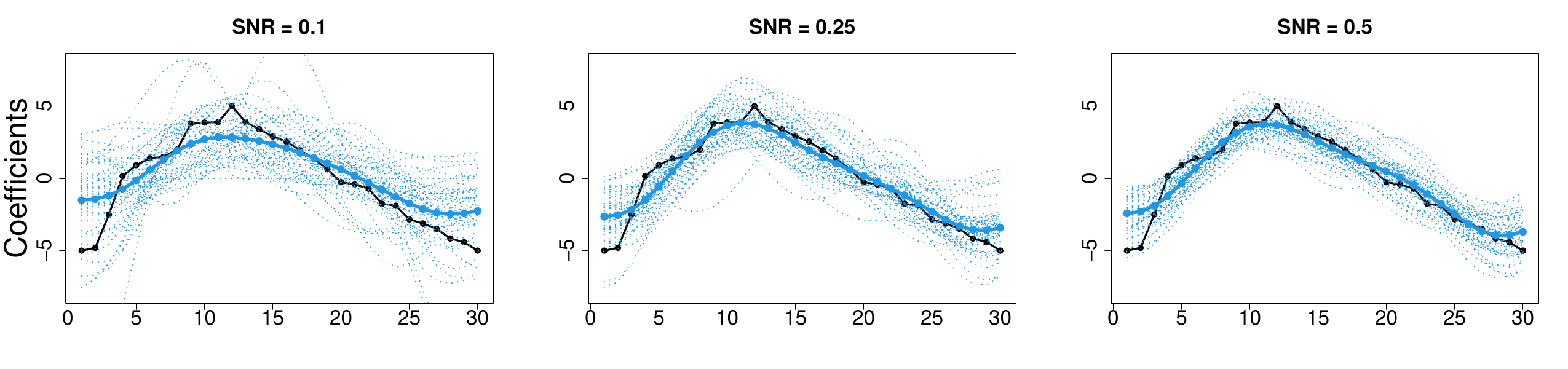}
  \end{minipage}
  \caption{Distributed lag curves of the 1st exposure in \textbf{Model A} with normal errors and different SNRs. Black lines are the true coefficients. We fix $n=750$ and vary SNRs. First row: coefficient estimates of \texttt{Uni} (green dotted) across 50 replications and the average (green solid) over those replications. 
  Second row: coefficient estimates of \texttt{Concave} (orange dotted) across 50 replications and the average (orange solid) over those replications. Third row: coefficient estimates of \texttt{FPCA-R} (blue dotted) across 50 replications and the average (blue solid) over those replications.}
  \label{fig::a6}
\end{figure}

\subsection{Results}
In Figure \ref{fig::new}, we present the estimation errors under the three data generating models with normal error distributions, varying sample sizes, and SNRs. The results with $t$-distribution errors are provided in the supplementary materials. The results demonstrate that our proposed estimators, \texttt{Uni} and \texttt{Concave}, perform favorably compared to the competitors across most scenarios, exhibiting smaller estimation errors.  While \texttt{FPCA-R} shows comparable or even better performance than \texttt{Uni} when $n=500$ and $\text{SNR}=0.1$, \texttt{Uni} consistently outperforms \texttt{FPCA-R} with larger sample sizes and higher SNRs. Similarly, \texttt{Concave} exhibits robust performance, maintaining lower estimation errors across all settings, and often outperforming other methods, especially in high SNR and large sample size scenarios.

Without imposing smoothness conditions on the distributed lag curves, \texttt{EN} and \texttt{Ridge} generally exhibit the worst performance in most settings. The sparsity structure in Model B does not provide a notable advantage to \texttt{EN}, which explicitly explores the sparsity in the coefficients. The performance gap indicates the importance of incorporating smoothness constraints in modeling distributed lag effects. The performance of \texttt{FPCA} is not as good as \texttt{FPCA-R}, suggesting that the ridge penalty is necessary for \texttt{FPCA} to achieve better performance.

In Figure \ref{fig::a6}, we illustrate the estimates of distributed lag curves for the first exposure in Model A, under normal errors and varying Signal-to-Noise Ratios (SNRs), as produced by \texttt{Uni}, \texttt{Concave} and \texttt{FPCA-R}. The results corresponding to Model B and Model C are provided in the supplementary materials. Each parameter setup is represented through plots that include both individual estimates from 50 replications and the corresponding average. While \texttt{FPCA-R} yields smooth estimates, it often falls short in capturing the unimodal characteristic inherent to the true distributed lag curves, with its average curve showing notable deviations at the curve's ends. In contrast, \texttt{Uni} and \texttt{Concave} reliably capture the unimodal shape of the true distributed lag curves, with their average estimate more accurately reflecting the actual curve.  

\section{Application to the Colorado birth cohort data}

\subsection{Data overview}
Finally, we use our method to analyze the Colorado birth cohort data \citep{mork2023estimating}. This dataset includes vital statistics records for births in Colorado, USA, with estimated conception dates between 2007 and 2015. The primary outcome of interest is the birth weight for gestational age z-score (BWGAZ). The dataset focuses on exposures to a mixture of environmental pollutants and temperature, which were measured at a high temporal resolution throughout the gestational period. Specifically, the exposures of interest include particulate matter smaller than 2.5 microns in diameter ($\text{PM}_{2.5}$), nitrogen dioxide ($\text{NO}_2$), sulfur dioxide ($\text{SO}_2$), carbon monoxide ($\text{CO}$), and temperature. These exposures were averaged over each week of gestation based on the mother's census tract of residence at delivery. For our analysis, we restricted the data to singleton, full-term births ($\ge$ 37 weeks) in the Denver metropolitan area, where exposure data is more accurately estimated. The final dataset consists of $n = 195,701$ births with complete covariate and exposure information. Covariates controlled for in the analysis include maternal age, weight, height, income, education, marital status, prenatal care habits, smoking habits, and race/ethnicity. Additionally, categorical variables for the year and month of conception, census tract elevation, and a county-specific intercept were included to adjust for potential confounders.

\subsection{Estimates of distributed lag curves and critical windows}

\begin{figure}[H]
  \centering
  \includegraphics[width=\textwidth]{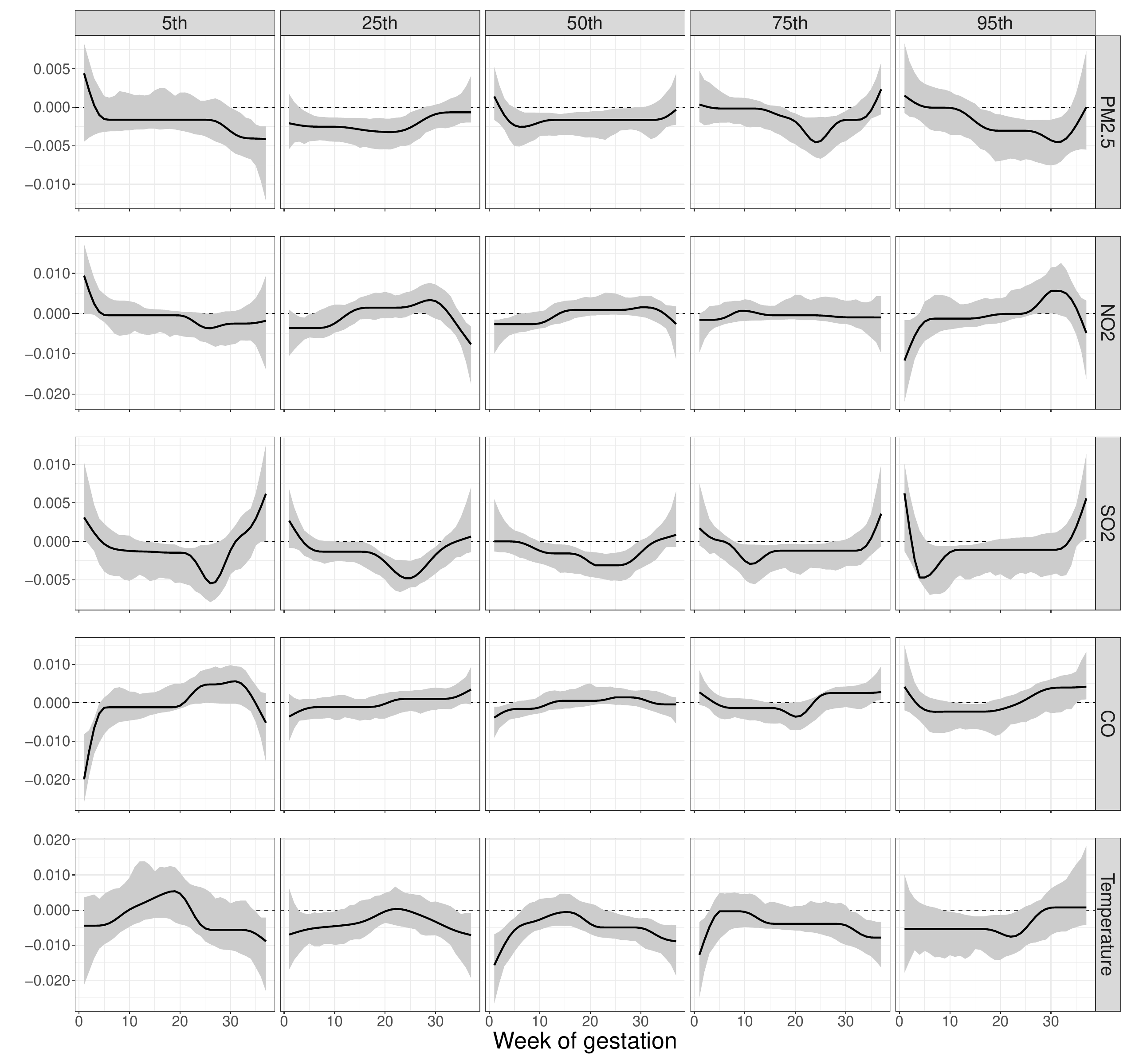}
  \vspace{-0.8cm}
  \caption{Distributed lag functions (solid line) for each exposure (rows) at different quantiles (columns) with 95\% confidence intervals (gray area). }
  \label{fig::dlc}
\end{figure}

\begin{figure}[t]
      \centering
      \begin{subfigure}[b]{0.49\textwidth}
        \centering
        \includegraphics[width=\textwidth]{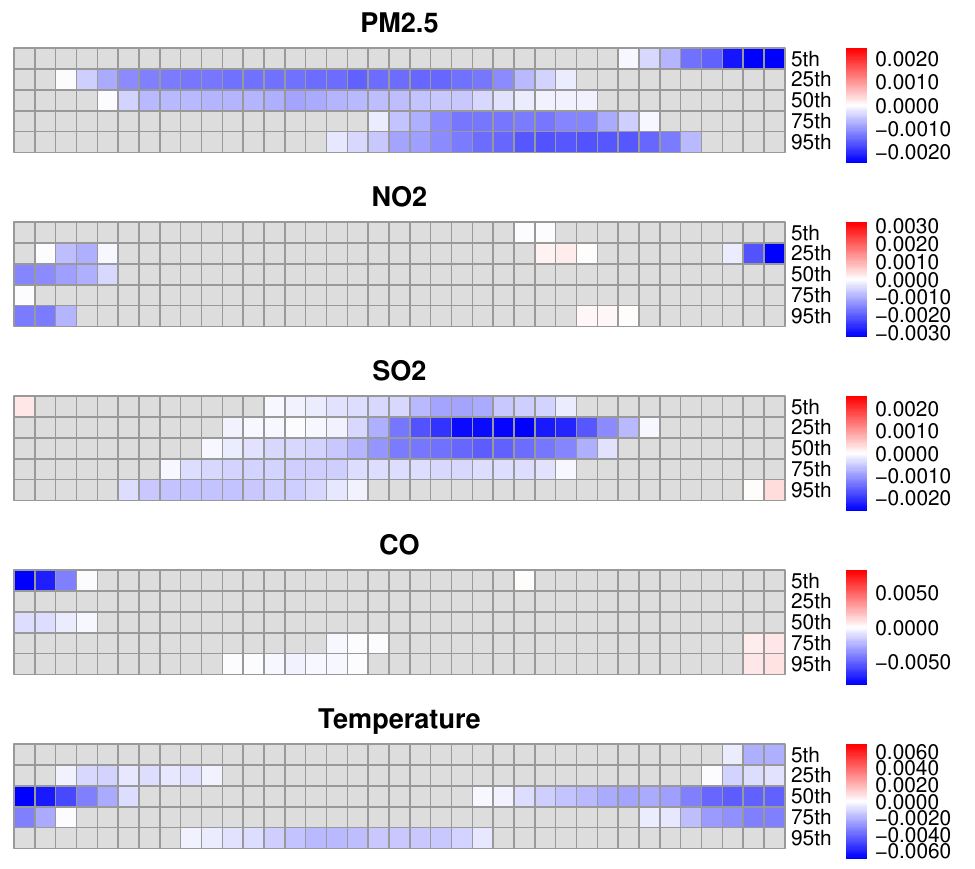}
        \caption{\texttt{Uni}}
        \label{fig::heatmap-concave}
      \end{subfigure}
      \hfill
      \begin{subfigure}[b]{0.49\textwidth}
        \centering
        \includegraphics[width=\textwidth]{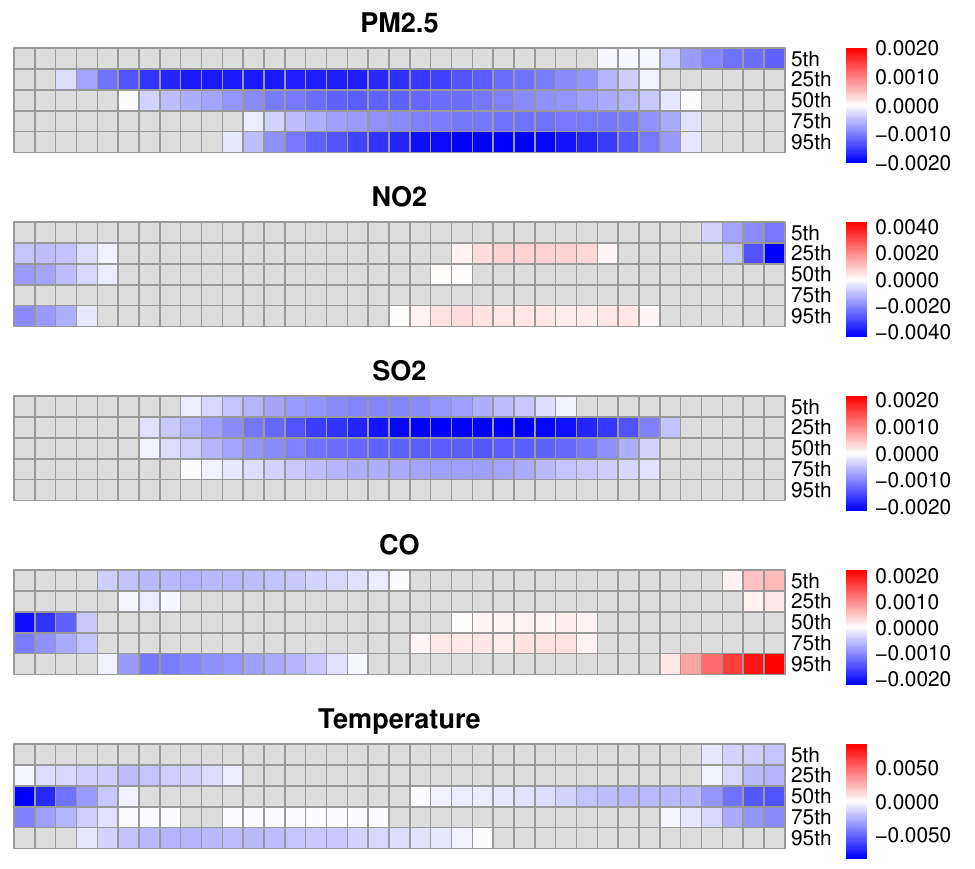}
        \caption{\texttt{Concave}}
        \label{fig::heatmap-concave-uni}
      \end{subfigure}
      \caption{Critical windows of different exposures identified by \texttt{Uni} and \texttt{Concave} across different quantiles. Grey areas represent the 95\% confidence intervals that include 0. Darker colors represent confidence intervals farther from 0.}
      \label{fig::CWs}
    \end{figure}

The estimates and corresponding pointwise 95\% confidence intervals for the coefficient estimates from the nearly unimodal estimator can be found in Figure \ref{fig::dlc}. Results were similar for the nearly concave estimator, and these have been relegated to the Appendix for brevity. The critical windows identified for both the nearly unimodal and nearly concave estimator can similarly be found in Figure \ref{fig::CWs}. A number of important findings about the adverse effects of environmental pollutants on gestational birth weight are evident from the results. First, there is a clear adverse effect of PM$_{2.5}$ on gestational birth weight indicating that exposure to increased levels of particular matter is associated with reductions in birth weight. Interestingly, there is some variability across quantiles in the time periods identified as critical windows for PM$_{2.5}$. Exposures in the latest time periods are estimated to be the most important for the 5th quantile, while time periods in the center of the pregnancy are most impactful on other quantiles, though there is some evidence that later time periods are most impactful for the upper quantiles as well. 

This ability to identify unique critical windows and effects for different quantiles highlights the important gains in inference that can be obtained from our estimator. Birth weight is an important predictor of many future health outcomes. However, effects on the extreme quantiles, as we show here with the 5th quantile, are clinically more important than a small shift in population mean.

In terms of the other exposures, only $\text{SO}_2$ and temperature have clear effects on gestational birth weight. The effect of $\text{SO}_2$ is focused in the center of the gestational period, while the effects of temperature are estimated to be most pronounced in the first and third trimesters of the pregnancy. Another interesting finding is found in the estimates and critical windows for carbon monoxide. Previous analyses of these data suggested a positive effect of CO during one part of the gestation period, and a negative effect in a different time period  \citep{mork2023estimating,antonelli2024multiple}. This is counter-intuitive because (i) we don't expect CO to be beneficial for children's health, and (ii) it is unexpected for the exposures to have one direction of association at one time period, and a different direction in another. Our nearly unimodal penalty precludes this from happening, showing the benefits in interpretability and scientific plausibility of findings that can be obtained when incorporating reasonable shape constraints into estimation. 

\section{Discussion}

In this paper we developed a novel penalized estimator for estimating quantile distributed lag functions when exposures are measured repeatedly over time, which allows for the incorporation of smoothness and shape constraint penalties that can improve estimation. We developed an efficient algorithm for computing the solution to the proposed nearly unimodal or nearly concave penalized objective functions, and showed in simulations that it performs well against feasible alternatives. Estimating distributed lag functions is a critically important problem in environmental health studies, particularly those involving maternal and children's health, and the ability to estimate conditional quantile distributed lags in a flexible manner without strong parametric assumptions on the distributed lag functions helps expand the set of questions that can be answered in such studies. We showed the benefits of the proposed approach in a large-scale study of births in Colorado, where the proposed approach identified critical windows of susceptibility for multiple exposures, while also providing more interpretable and plausible results than prior research in this cohort. 

There are a number of distinct directions that could add to the utility of the proposed framework for estimating quantile distributed lag functions. While we have incorporated smoothness and shape constraints through penalties on the estimated coefficients, we do not currently enforce sparsity, which could be beneficial in high-dimensional settings or when exposure selection is desired. While incorporating lasso or group lasso penalties could potentially be a useful direction to pursue, this would come with computational challenges from the introduction of additional tuning parameters. Another direction of relevance in environmental epidemiology is to pursue bivariate distributed lag functions that allow for interactions between two exposures at different time points \citep{chen2019distributed}. The extension of smoothness penalties is relatively straightforward in this direction, but the incorporation of shape constraints is not trivial, and could improve estimation by incorporating known restrictions for two-dimensional distributed lag surfaces.

\section*{Acknowledgements}

These data were supplied by the Center for Health and Environmental Data Vital Statistics Program of the Colorado Department of Public Health and Environment, which specifically disclaims responsibility for any analyses, interpretations, or conclusions it has not provided. The authors thank Ben Sherwood for feedback on an earlier version of this article.

\bibliography{base}

\begin{thebibliography}{}

\bibitem[Antonelli et~al., 2024]{antonelli2024multiple}
Antonelli, J., Wilson, A., and Coull, B.~A. (2024).
\newblock Multiple exposure distributed lag models with variable selection.
\newblock {\em Biostatistics}, 25(1):1--19.

\bibitem[Bosetti et~al., 2010]{bosetti2010ambient}
Bosetti, C., Nieuwenhuijsen, M.~J., Gallus, S., Cipriani, S., La~Vecchia, C.,
  and Parazzini, F. (2010).
\newblock Ambient particulate matter and preterm birth or birth weight: a
  review of the literature.
\newblock {\em Archives of Toxicology}, 84:447--460.

\bibitem[Boyd et~al., 2011]{boyd2011distributed}
Boyd, S., Parikh, N., Chu, E., Peleato, B., and Eckstein, J. (2011).
\newblock Distributed optimization and statistical learning via the alternating
  direction method of multipliers.
\newblock {\em Foundations and Trends in Machine Learning}, 3(1):1--122.

\bibitem[Brantley et~al., 2020]{brantley2020baseline}
Brantley, H.~L., Guinness, J., and Chi, E.~C. (2020).
\newblock Baseline drift estimation for air quality data using quantile trend
  filtering.
\newblock {\em Annals of Applied Statistics}, 14(2).

\bibitem[Chen et~al., 2018]{chen2018robust}
Chen, Y.-H., Mukherjee, B., Adar, S.~D., Berrocal, V.~J., and Coull, B.~A.
  (2018).
\newblock Robust distributed lag models using data adaptive shrinkage.
\newblock {\em Biostatistics}, 19(4):461--478.

\bibitem[Chen et~al., 2019]{chen2019distributed}
Chen, Y.-H., Mukherjee, B., and Berrocal, V.~J. (2019).
\newblock Distributed lag interaction models with two pollutants.
\newblock {\em Journal of the Royal Statistical Society Series C: Applied
  Statistics}, 68(1):79--97.

\bibitem[Deng and Yin, 2016]{deng2016global}
Deng, W. and Yin, W. (2016).
\newblock On the global and linear convergence of the generalized alternating
  direction method of multipliers.
\newblock {\em Journal of Scientific Computing}, 66:889--916.

\bibitem[Feng et~al., 2011]{feng2011wild}
Feng, X., He, X., and Hu, J. (2011).
\newblock Wild bootstrap for quantile regression.
\newblock {\em Biometrika}, 98(4):995--999.

\bibitem[Gasparrini, 2014]{gasparrini2014modeling}
Gasparrini, A. (2014).
\newblock Modeling exposure--lag--response associations with distributed lag
  non-linear models.
\newblock {\em Statistics in Medicine}, 33(5):881--899.

\bibitem[Gasparrini et~al., 2017]{gasparrini2017penalized}
Gasparrini, A., Scheipl, F., Armstrong, B., and Kenward, M.~G. (2017).
\newblock A penalized framework for distributed lag non-linear models.
\newblock {\em Biometrics}, 73(3):938--948.

\bibitem[Ghosal et~al., 2023]{ghosal2023shape}
Ghosal, R., Ghosh, S., Urbanek, J., Schrack, J.~A., and Zipunnikov, V. (2023).
\newblock Shape-constrained estimation in functional regression with bernstein
  polynomials.
\newblock {\em Computational Statistics and Data Analysis}, 178:107614.

\bibitem[Goldsmith et~al., 2023]{Goldsmith2023}
Goldsmith, J., Scheipl, F., Huang, L., Wrobel, J., Di, C., Gellar, J.,
  Harezlak, J., McLean, M.~W., Swihart, B., Xiao, L., Crainiceanu, C., and
  Reiss, P.~T. (2023).
\newblock {\em refund: Regression with Functional Data}.
\newblock R package version 0.1-34.

\bibitem[Gu et~al., 2018]{gu2018admm}
Gu, Y., Fan, J., Kong, L., Ma, S., and Zou, H. (2018).
\newblock Admm for high-dimensional sparse penalized quantile regression.
\newblock {\em Technometrics}, 60(3):319--331.

\bibitem[Jacobs et~al., 2017]{jacobs2017association}
Jacobs, M., Zhang, G., Chen, S., Mullins, B., Bell, M., Jin, L., Guo, Y.,
  Huxley, R., and Pereira, G. (2017).
\newblock The association between ambient air pollution and selected adverse
  pregnancy outcomes in china: a systematic review.
\newblock {\em Science of the Total Environment}, 579:1179--1192.

\bibitem[Joubert et~al., 2022]{joubert2022powering}
Joubert, B.~R., Kioumourtzoglou, M.-A., Chamberlain, T., Chen, H.~Y., Gennings,
  C., Turyk, M.~E., Miranda, M.~L., Webster, T.~F., Ensor, K.~B., and Dunson,
  D.~B. (2022).
\newblock Powering research through innovative methods for mixtures in
  epidemiology (prime) program: novel and expanded statistical methods.
\newblock {\em International Journal of Environmental Research and Public
  Health}, 19(3):1378.

\bibitem[Koenker and Bassett~Jr, 1978]{koenker1978regression}
Koenker, R. and Bassett~Jr, G. (1978).
\newblock Regression quantiles.
\newblock {\em Econometrica}, pages 33--50.

\bibitem[Koenker et~al., 2017]{regression2017handbook}
Koenker, R., Chernozhukov, V., He, X., and Peng, L. (2017).
\newblock {\em Handbook of Quantile Regression}.
\newblock CRC Press: Boca Raton, FL, USA.

\bibitem[Lange, 2016]{lange2016mm}
Lange, K. (2016).
\newblock {\em MM optimization algorithms}.
\newblock SIAM.

\bibitem[Liu, 1988]{liu1988bootstrap}
Liu, R.~Y. (1988).
\newblock Bootstrap procedures under some non-{IID} models.
\newblock {\em Annals of Statistics}, 16(4):1696--1708.

\bibitem[Mork and Wilson, 2023]{mork2023estimating}
Mork, D. and Wilson, A. (2023).
\newblock Estimating perinatal critical windows of susceptibility to
  environmental mixtures via structured bayesian regression tree pairs.
\newblock {\em Biometrics}, 79(1):449--461.

\bibitem[Obermeier et~al., 2015]{obermeier2015flexible}
Obermeier, V., Scheipl, F., Heumann, C., Wassermann, J., and K{\"u}chenhoff, H.
  (2015).
\newblock Flexible distributed lags for modelling earthquake data.
\newblock {\em Journal of the Royal Statistical Society Series C: Applied
  Statistics}, 64(2):395--412.

\bibitem[Schwartz, 2000]{schwartz2000distributed}
Schwartz, J. (2000).
\newblock The distributed lag between air pollution and daily deaths.
\newblock {\em Epidemiology}, 11(3):320--326.

\bibitem[Stieb et~al., 2012]{stieb2012ambient}
Stieb, D.~M., Chen, L., Eshoul, M., and Judek, S. (2012).
\newblock Ambient air pollution, birth weight and preterm birth: a systematic
  review and meta-analysis.
\newblock {\em Environmental Research}, 117:100--111.

\bibitem[Tan et~al., 2022]{tan2022high}
Tan, K.~M., Wang, L., and Zhou, W.-X. (2022).
\newblock High-dimensional quantile regression: Convolution smoothing and
  concave regularization.
\newblock {\em Journal of the Royal Statistical Society Series B: Statistical
  Methodology}, 84(1):205--233.

\bibitem[Tibshirani, 2014]{tibshirani2014adaptive}
Tibshirani, R.~J. (2014).
\newblock Adaptive piecewise polynomial estimation via trend filtering.
\newblock {\em Annals of Statistics}, 42(1):285.

\bibitem[Tibshirani et~al., 2011]{tibshirani2011nearly}
Tibshirani, R.~J., Hoefling, H., and Tibshirani, R. (2011).
\newblock Nearly-isotonic regression.
\newblock {\em Technometrics}, 53(1):54--61.

\bibitem[Wang et~al., 2018]{wang2018wild}
Wang, L., Van~Keilegom, I., and Maidman, A. (2018).
\newblock Wild residual bootstrap inference for penalized quantile regression
  with heteroscedastic errors.
\newblock {\em Biometrika}, 105(4):859--872.

\bibitem[Wang et~al., 2023]{wang2023semiparametric}
Wang, Y., Ghassabian, A., Gu, B., Afanasyeva, Y., Li, Y., Trasande, L., and
  Liu, M. (2023).
\newblock Semiparametric distributed lag quantile regression for modeling
  time-dependent exposure mixtures.
\newblock {\em Biometrics}, 79(3):2619--2632.

\bibitem[Warren et~al., 2012]{warren2012spatial}
Warren, J., Fuentes, M., Herring, A., and Langlois, P. (2012).
\newblock Spatial-temporal modeling of the association between air pollution
  exposure and preterm birth: identifying critical windows of exposure.
\newblock {\em Biometrics}, 68(4):1157--1167.

\bibitem[Warren et~al., 2022]{warren2022critical}
Warren, J.~L., Chang, H.~H., Warren, L.~K., Strickland, M.~J., Darrow, L.~A.,
  and Mulholland, J.~A. (2022).
\newblock Critical window variable selection for mixtures: estimating the
  impact of multiple air pollutants on stillbirth.
\newblock {\em Annals of Applied Statistics}, 16(3):1633.

\bibitem[Wilson et~al., 2017a]{wilson2017bayesian}
Wilson, A., Chiu, Y.-H.~M., Hsu, H.-H.~L., Wright, R.~O., Wright, R.~J., and
  Coull, B.~A. (2017a).
\newblock Bayesian distributed lag interaction models to identify perinatal
  windows of vulnerability in children’s health.
\newblock {\em Biostatistics}, 18(3):537--552.

\bibitem[Wilson et~al., 2017b]{wilson2017potential}
Wilson, A., Chiu, Y.-H.~M., Hsu, H.-H.~L., Wright, R.~O., Wright, R.~J., and
  Coull, B.~A. (2017b).
\newblock Potential for bias when estimating critical windows for air pollution
  in children’s health.
\newblock {\em American Journal of Epidemiology}, 186(11):1281--1289.

\bibitem[Wilson et~al., 2022]{wilson2022kernel}
Wilson, A., Hsu, H.-H.~L., Chiu, Y.-H.~M., Wright, R.~O., Wright, R.~J., and
  Coull, B.~A. (2022).
\newblock Kernel machine and distributed lag models for assessing windows of
  susceptibility to environmental mixtures in children’s health studies.
\newblock {\em Annals of Applied Statistics}, 16(2):1090.

\bibitem[Wright, 2017]{wright2017environment}
Wright, R.~O. (2017).
\newblock Environment, susceptibility windows, development, and child health.
\newblock {\em Current Opinion in Pediatrics}, 29(2):211--217.

\bibitem[Wu, 1986]{wu1986jackknife}
Wu, C.-F.~J. (1986).
\newblock Jackknife, bootstrap and other resampling methods in regression
  analysis.
\newblock {\em Annals of Statistics}, 14(4):1261--1295.

\bibitem[Xiao et~al., 2016]{xiao2016fast}
Xiao, L., Zipunnikov, V., Ruppert, D., and Crainiceanu, C. (2016).
\newblock Fast covariance estimation for high-dimensional functional data.
\newblock {\em Statistics and Computing}, 26:409--421.

\bibitem[Yu et~al., 2003]{yu2003quantile}
Yu, K., Lu, Z., and Stander, J. (2003).
\newblock Quantile regression: applications and current research areas.
\newblock {\em Journal of the Royal Statistical Society Series D: The
  Statistician}, 52(3):331--350.

\bibitem[Zanobetti et~al., 2000]{zanobetti2000generalized}
Zanobetti, A., Wand, M.~P., Schwartz, J., and Ryan, L.~M. (2000).
\newblock Generalized additive distributed lag models: quantifying mortality
  displacement.
\newblock {\em Biostatistics}, 1(3):279--292.

\end{thebibliography}
\newpage

\end{document}